\documentclass[showpacs]{revtex4}
\usepackage{epsfig,amssymb}
\newcommand{\anti}[1]{\overline{#1}}

\begin{document}

%\title{\hfill {\footnotesize FZJ--IKP(TH)--2005--20, HISKP-TH-05-29}\\
% The J\"ulich hyperon-nucleon model revisited}

\title{The J\"ulich hyperon-nucleon model revisited}

\author{J. Haidenbauer$^1$ and Ulf-G. Mei{\ss}ner$^{1,2}$}

\affiliation{
$^1$Institut f\"ur Kernphysik (Theorie), Forschungszentrum J\"ulich,
D-52425 J\"ulich, Germany,\\
$^2$Helmholtz-Institut f\"ur Strahlen- und Kernphysik (Theorie), 
Universit\"at Bonn, Nu\ss allee 14-16, D-53115 Bonn, Germany 
}

\begin{abstract}
A one-boson-exchange model for the hyperon-nucleon ($\Lambda N$, $\Sigma N$) interaction
is presented. The model incorporates the standard one boson exchanges of the lowest 
pseudoscalar and vector meson multiplets with coupling
constants fixed by SU(6) flavor symmetry relations. As the main new feature of 
the model, the contributions in the scalar--isoscalar ($\sigma$) 
and vector--isovector ($\rho$) exchange channels are now constrained by a
microscopic model of correlated $\pi\pi$ and $K \bar K$ exchange. 
Additional short-ranged ingredients of the model in the scalar--isovector ($a_0$) 
and scalar--isospin-1/2 ($\kappa$) channels are likewise viewed as arising
from meson-meson correlations but are treated phenomenologically. 
With this model a satisfactory reproduction of the available 
hyperon-nucleon data can be achieved. 
\end{abstract}
\pacs{13.75.Ev, 12.39.Pn, 21.30.-x, 21.80.+a} 

\maketitle

%%%%%%%%%%%%%%%%%%%%%%%%%%%%%%%%%%%%%%%%%%%%%%%%%%%%%%%%%%%%%%%%%%%%%%%
\section{Introduction}

The role of strangeness in low and medium energy nuclear physics is currently
of considerable interest, as it has the potential to deepen our
understanding of the relevant strong interaction mechanisms in
the non-perturbative regime of QCD.
For example, the system of a strange baryon (hyperon $Y$) and a
nucleon ($N$) is in principle an ideal testing ground for investigating
the importance of SU(3)$_{flavor}$ symmetry in hadronic interactions.
Existing meson exchange models of the $YN$ force usually assume SU(3)
symmetry for the hadronic coupling constants, and in some
cases \cite{Holz,Reu} even the SU(6) symmetry of the quark model.
The symmetry requirements provide relations between couplings of
mesons of a given multiplet to the baryon current, which greatly
reduce the number of free model parameters.
Specifically, coupling constants at the strange vertices are then
connected to nucleon-nucleon-meson coupling constants, which in
turn are constrained by the wealth of empirical information on $NN$
scattering.  Essentially all these $YN$ interaction models can reproduce the 
existing $YN$ scattering
data, so that at present the assumption of SU(3) symmetry for the
coupling constants cannot be ruled out by experiment.

One should note, however, that the various models differ dramatically
in the treatment of the scalar-isoscalar meson sector, which describes
the baryon-baryon interaction at intermediate ranges.
For example, the Nijmegen group \cite{NijII,NijIII,NijIV,NijV} views this
interaction as being generated by genuine scalar meson exchange.
In their model D \cite{NijII} an $\epsilon(760)$ is exchanged as an
SU(3)$_{\it flavor}$ singlet.
In models F~\cite{NijIII}, NSC~\cite{NijIV}, and NSC97 \cite{NijV} a scalar 
SU(3) nonet is exchanged --- namely, two isospin-0 mesons (besides the $\epsilon(760)$, 
the $\epsilon '(1250)$ in model F and $S^*(975)$ ($f_0(980)$) in model NSC (NSC97)), 
an isospin-1 meson ($\delta$ or $a_0(980)$) and an isospin-1/2 strange meson $\kappa$
with a mass of 1000 MeV. 
A genuine scalar SU(3) nonet is also present in the so-called Ehime 
potential \cite{Ehime}, where besides the $S^*(975)$ and $\delta$ (or $a_0(980)$)
the $f_0(1581)$ and the $K_0^*(1429)$ are included. In additon the model incorporates
two effective scalar-meson exchanges, $\sigma (484)$ and $\kappa (839)$, that stand
for $(\pi\pi)_{I=0}$ and $(K\pi)_{I=1/2}$ correlations but are treated 
phenomenologically. The T\"ubingen model, on the other hand, which is
essentially a constituent quark model supplemented by $\pi$ and
$\sigma$ exchange at intermediate and short ranges, treats the
$\sigma$ meson as an SU(3) singlet with a mass of 520 MeV \cite{Tueb}
or 675 MeV \cite{Zhang1}, respectively. Finally, in the quark-models of 
Zhang et al. \cite{Zhang2} and Fujiwara et al. \cite{Fujiwara} 
a scalar SU(3) nonet is exchanged, though in this case between 
quarks and not between the
baryons. 

\begin{figure}[h]
\vskip 4cm
\includegraphics{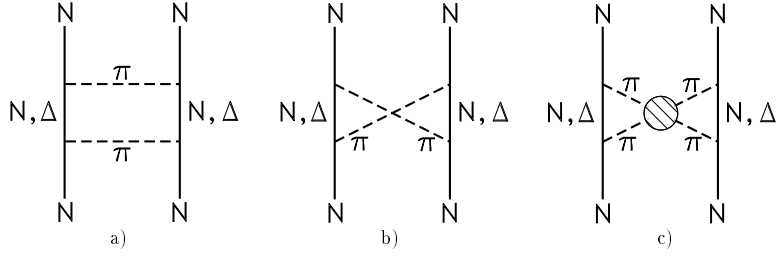} 
\caption{Two-pion exchange in the $NN$ interaction: (a) uncorrelated
        iterative and (b) crossed boxes, (c) correlated two-pion exchange.}
\label{fig1}
\end{figure}

In the (full) Bonn $NN$ potential~\cite{MHE} the intermediate range
attraction is provided by uncorrelated and correlated $\pi\pi$ exchange
processes (Figs.~\ref{fig1}(a)--(b) and Fig.~\ref{fig1}(c), respectively),
with $NN$, $N\Delta$ and $\Delta\Delta$ intermediate states.
{}From earlier studies of the $\pi\pi$ interaction it is known that
$\pi\pi$ correlations are important mainly in the scalar-isoscalar
and vector-isovector channels.
In one-boson-exchange (OBE) potentials these are included effectively
via exchange of sharp mass $\sigma$ and $\rho$ mesons.
One disadvantage of such a simplified treatment is that this
parameterization cannot be transferred into the hyperon sector
in a well defined manner.
Therefore in the earlier $YN$ interaction models of the J\"ulich 
group~\cite{Holz}, which start from the Bonn $NN$ potential,
the coupling constants of the fictitious $\sigma$ meson at the
strange vertices ($\Lambda\Lambda\sigma$, $\Sigma\Sigma\sigma$)
are free parameters --- a rather unsatisfactory feature of the
models.
This is especially true for the extension to the strangeness $S=-2$
channels, interest in which initiated with the prediction of the
H-dibaryon by Jaffe~\cite{Jaffe}.

These problems can be overcome by an explicit evaluation of correlated
$\pi\pi$ exchange in the various baryon-baryon channels.
A corresponding calculation was initially done only for the $NN$ case
(Fig.~\ref{fig1}(c)) in Ref. \cite{Kim},
but was extended in a recent paper \cite{REUBER} 
by the J\"ulich group so that now a full and consistent 
microscopic derivation of correlated $\pi\pi$ exchange in various
baryon-baryon ($BB'$) channels with strangeness $S=0, -1$ and $-2$
is available. 
The starting point was a field theoretical model for both the
$N\anti{N}\to\pi\pi$ Born amplitudes and the $\pi\pi$ and $K\anti{K}$
elastic scattering~\cite{Lohse,Janssen,Schutz}. Thus, 
the $K\anti{K}$ channel is treated on an equal footing with the
$\pi\pi$ channel in order to reliably determine the influence of
$K\anti{K}$ correlations in the relevant $t$-channels.
Then, with the help of unitarity and dispersion relations the amplitude
for the correlated $\pi\pi$ exchange in the $NN$ channel but also
for the $YN$ and $YY$ systems were computed. Thus, within
this approach one can replace the phenomenological $\sigma$
and $\rho$ exchanges in the Bonn $NN$ \cite{MHE} and J\"ulich $YN$
\cite{Holz} models by correlated processes, i.e. eliminate undetermined
parameters such as the $BB'\sigma$ coupling constants.

In the present paper a new $YN$ model is presented that utilizes
this microscopic model of correlated $\pi\pi$ and $K\bar K$ exchange 
to fix the contributions in the scalar-isoscalar ($\sigma$) and 
vector-isovector ($\rho$) channels. The model incorporates also
the standard one boson exchange contributions of the lowest 
pseudoscalar and vector meson multiplets with coupling constants determined 
by SU(6) symmetry relations. Assuming the SU(6) symmetry means that 
also the so-called $F/(F+D)$ ratios are fixed. In addition, there are further 
new ingrediations as compared to the original J\"ulich $YN$ model \cite{Holz}.
First of all, the contribution from the $a_0(980)$ meson is taken into
account. Secondly, we consider the exchange of a strange scalar meson, the
$\kappa$, with mass $\sim 1000$~MeV.
Let us emphasize, however, that in analogy with the $\sigma$ meson these particles 
are likewise not viewed as being members of a scalar meson SU(3) multiplet, but 
rather as representations of strong meson-meson correlations in the 
scalar--isospin-1/2 ($\pi K$) \cite{Lohse} and
scalar--isovector ($\pi\eta$--$K\bar K$) \cite{Janssen} channels 
respectively.
In principle, their contributions can also be evaluated along the lines
of Ref.~\cite{REUBER}, however, for simplicity in the present model they
are effectively parameterized by one-boson-exchange diagrams with the
appropriate quantum numbers assuming the coupling constants to be free
parameters.

In the next two sections we describe the principal steps of the 
derivation of correlated $\pi\pi$ and $K\anti{K}$ exchange potentials for
the baryon--baryon amplitudes in the $\sigma$ and $\rho$ channels.
In particular, in Sect. 2 we give a short outline of the microscopic model for 
the required $B\anti{B'}\to\pi\pi,\,K\anti{K}$ amplitudes.
The derivation of the potentials themselves is indicated in Section 3. 
Furthermore, we introduce and discuss the parameterization of correlated 
$\pi\pi$ and $K\anti{K}$ exchange potentials by an effective $\sigma$ 
and $\rho$ exchange for the $YN$ channels. These effective 
parameterizations are then adopted for the construction of the new
$YN$ model. 
In Sect. 4 the other ingredients of our $YN$ model are introduced. Specifically, 
we comment on the employed strategy for fixing the parameters of the model.
Then we present and discuss numerical results of the model for $YN$ 
scattering observables, phase shifts and effective range parameters. 
Finally, some concluding remarks are made in Section 5.

%%%%%%%%%%%%%%%%%%%%%%%%%%%%%%%%%%%%%%%%%%%%%%%%%%%%%%%%%%%%%%%%%%%%%%%%%%
\section{Model for correlated $2\pi$ exchange}

Based on a $\pi\pi - K\bar K$ amplitude 
the evaluation of diagrams such as in
Fig.~\ref{fig1}(c) for any $BB'$ system can be done in two steps.
Firstly the $N\anti{N} \ (\Lambda \anti{\Lambda}, \ \Sigma \anti{\Sigma},
\ {\rm etc}.) \rightarrow 2\pi, K\bar K$ amplitudes 
are determined in the pseudophysical region ($t \leq 4 m^2_\pi$) 
and then dispersion theory and unitarity are applied to connect those
amplitudes with the corresponding physical amplitudes in the various
baryon-baryon channels. 

Figure \ref{fig5} shows a graphic representation of our dynamical model
for correlated $2\pi - K\bar K$ exchange.
Here $B\anti{B'}$ stands for $N\anti{N}$, $\Lambda \anti{ \Lambda}$,
$\Lambda \anti{ \Sigma}$/$\Sigma \anti{ \Lambda}$
or $\Sigma \anti{ \Sigma}$. Formally the amplitudes for the processes 
$B\anti{B'} \rightarrow \alpha$ (with $\alpha = \pi\pi$, $K\anti{K}$) are obtained 
from solving the scattering equation 
\begin{equation}
T_{B,\overline{B'} \to \alpha} = 
V_{B,\overline{B'} \to \alpha} + 
\sum_{\beta = \pi\pi, K\bar K} T_{\alpha,\beta} \,
G_{\beta} \, V_{B,\overline{B'} \to \beta} \ .
\label{DWBA}
\end{equation}
Here $T_{\alpha,\beta}$ is the $\pi\pi-K\bar K$ (coupled-channel)
reaction amplitude, $V_{B,\overline{B'} \to \beta}$ the 
$B\overline{B'} \to \pi\pi,K\bar K$ transition Born amplitude and
$G_{\beta}$ the free ($\pi\pi$ or $K\bar K$) Green's function 
The first two quantities are the basic ingredients of the model. 
These amplitudes have to be known in the so-called pseudophysical region, 
i.e. for energies below the $B\anti{B'}$ threshold. While for $N\anti{N} \rightarrow 
\pi\pi$ the corresponding amplitudes can be derived from empirical 
information on $\pi N$ and $\pi\pi$ scattering via an analytic continuation,
this is not possible for the transitions $Y\anti{Y'} \rightarrow \pi\pi, K\bar K$.
Thus, a microscopic model for the $B\anti{B'} \rightarrow \pi\pi$,$K\anti{K}$
is a pre-requisite for the evaluation of the correlated $\pi\pi$ and $K\bar K$
exchange in the $YN$ and $YY$ channels. Such a model was constructed by 
Reuber et al. in Ref. \cite{REUBER}. A main feature of this model 
is the completely consistent treatment of its two components, namely
the $B\anti{B'} \rightarrow \pi\pi, K\anti{K}$ Born amplitudes and the
$\pi\pi$-$K\anti{K}$ correlations with respect to the $\sigma$- and $\rho$-channels.
Both components are derived in field theory from an ansatz for the hadronic
Lagrangians \cite{REUBER}. The considered contributions are briefly described 
in the subsections below. The subsequent evaluation of the baryon-baryon interaction
via dispersion theory and unitarity is summarized in the next section. 
 
\subsection{The $\pi\pi - K\bar K$ amplitudes} 

The dynamical model used for the $\pi\pi - K\bar K$ amplitudes is 
derived within the meson exchange framework and involves the 
$\pi\pi$ and $K\anti{K}$ coupled channels \cite{Lohse,Janssen,Schutz}.
The driving terms for the diagonal interactions consist of ($t$-channel) exchange
diagrams ($\rho$ and $\rho$, $\omega$, $\phi$, respectively) and ($s$-channel) 
pole diagrams with $\epsilon \equiv f_0(1440)$, $\rho \equiv \rho (770)$ and
$f_2 \equiv f_2(1274)$ intermediate states.
The coupling $\pi\pi \rightarrow K\anti{K}$ is provided by $K^*(892)$
exchange. The corresponding diagrams are shown in Fig. \ref{pipi}. 
The potentials derived from those diagrams are iterated in a coupled-channel
Lippmann-Schwinger-type scattering equation. The free parameters of the $\pi\pi - K\bar K$ 
model were adjusted to the empirical $\pi\pi$ phase shifts and inelasticities. 
For details on the model and a comparision of the resulting $\pi\pi$
phase shifts with experimental values we refer the reader to Refs. 
\cite{Janssen,Schutz}.

\subsection{The $B\anti{B} \rightarrow 2\pi,K\bar K$ Born amplitudes} 

The Born amplitudes for the transition $B\anti{B} \rightarrow \alpha$ with
$\alpha = \pi\pi,K\bar K$ are built up from an ($s$-channel) 
$\rho$-pole diagram and all possible diagrams involving the exchange of baryons out of the 
$J^P = {1\over 2}^+$ octet or the $J^P = {3\over 2}^+$ decuplet \cite{REUBER}. 
For illustration we show in Fig. \ref{Born} those diagrams that contribute to
the transition amplitude for $\Sigma\anti{\Sigma} \rightarrow 2\pi,K\bar K$.
 
In the construction of the model the number of free parameters has been kept to
a minimum. Specifically, 
the coupling constants at the various vertices involving the pseudoscalar mesons
were fixed by SU(6) symmetry relations. As far as the $\rho$-pole diagram is concerned 
the (bare) coupling constants and form factors at the $\pi\pi \rho^{(0)}$ and
$K\bar K \rho^{(0)}$ vertices were already determined in the model for the
$\pi\pi-K\bar K$ interaction \cite{Schutz} and were taken over from there. Then, 
assuming that the bare $\rho$-meson couples universally to the isospin current all 
vector couplings $g^{(0)}_{BB'\rho}$ to the baryonic vertices were fixed as well. 
For the tensor couplings $f^{(0)}_{BB'\rho}$ again SU(6) symmetry relations were
applied.

The four remaining free parameters (the tensor coupling constant $f^{(0)}_{NN\rho}$,
the parameter $x_\Delta$ characterizing the strength of the off-shell 
part in the $\Delta N\pi$ Lagrangian\footnote{In a more modern language, it
can be shown that such off-shell parameters really correspond to low-energy
constants of non-propagating contact interactions, see e.g. \cite{BKMdelta}}, 
and form-factor parameters for the 
exchanged baryons for the octet and decuplet, respectively \cite{REUBER})
were fixed by adjusting the model predictions to the 
quasi-empirical information on the amplitudes $N\bar N \to \pi\pi$ obtained by 
H\"ohler at el.~\cite{Hoehler2} by analytically continuing the $\pi N$ 
and $\pi \pi$ scattering data.
Once this is done the model can be used to generate the amplitudes for 
any $B\bar B' \to \pi\pi, K\bar K$ channel. Though it should have become
clear already from the discussion above, we want to emphasize again that the
extrapolation of the model for the $N\bar N \to \pi\pi$ amplitudes to other
channels depends crucially on the assumption of SU(3) symmetry 
for the pseudo-scalar sector and it is based also on the hope that the
correct description of the quasiempirical $N\bar N \to \pi\pi$ amplitudes
guarantees a reasonable description of the other baryon-antibaryon channels,
for which no empirical information is available.

%%%%%%%%%%%%%%%%%%%%%%%%%%%%%%%%%%%%%%%%%%%%%%%%%%%%%%%%%%%%%%%%%%%%%%%
\section{Potential from correlated $\pi\pi$ and $K\anti{K}$ exchange}

Assuming analyticity for the amplitudes dispersion relations
can be formulated for the baryon-baryon amplitudes, which connect
physical amplitudes in the $s$-channel with singularities and
discontinuities of these amplitudes in the pseudophysical region of
the $t$-channel processes for the $J^P = 0^+$ ($\sigma$) and $1^-$ ($\rho$)
channel:
\begin{equation}
V^{(0^+,1^-)}_{B_1,B_2 \to B_1',B_2'}(t) \propto \int_{4m^2_\pi}^\infty
dt' 
{ {\rm Im} V^{(0^+,1^-)}_{B_1,\overline{B_1'} \to \overline{B_2},B_2'}(t') \over t'-t}, \ \  t < 0 .
\label{dispersion}
\end{equation}
Via unitarity relations the singularity structure of the baryon-baryon 
amplitudes for $\pi\pi$ and $K\anti{K}$ exchange are fixed by and 
can be written as products of the 
$B\anti{B'}\to\pi\pi,\,K\anti{K}$ amplitudes 
\begin{equation}
{\rm Im} V^{(0^+,1^-)}_{B_1,\overline{B_1'} \to \overline{B_2},B_2'}(t') \propto
\sum_{\alpha = \pi\pi, K\bar K} T^{*,(0^+,1^-)}_{B_1,\overline{B_1'} \to \alpha}
\, T^{(0^+,1^-)}_{\overline{B_2},B_2' \to \alpha}.
\label{unitarity}
\end{equation}
Thus, 
from the $B\anti{B'} \rightarrow 2\pi $ helicity amplitudes the
spectral functions can be calculated 
\begin{equation}
\rho^{(0^+,1^-)}_{B_1,B_2 \to B_1',B_2'}(t') \propto
\sum_{\alpha = \pi\pi, K\bar K} T^{*,(0^+,1^-)}_{B_1,\bar{B_1'} \to \alpha}
\, T^{(0^+,1^-)}_{\bar{B_2},B_2' \to \alpha} 
\label{spectral}
\end{equation}
which are then inserted into dispersion integrals to
obtain the (on-shell) baryon-baryon interaction:
\begin{equation}
V^{(0^+,1^-)}_{B_1,B_2 \to B_1',B_2'}(t) \propto \int_{4m^2_\pi}^\infty
dt' 
{\rho^{(0^+,1^-)}_{B_1,B_2 \to B_1',B_2'}(t') \over t'-t}, \ \  t < 0 .
\label{potential}
\end{equation}
The underlying formalism is quite involved and has been outlined
in detail already in Ref.~\cite{REUBER}. Thus, we refrain from 
repeating it here. Rather we want to provide only some more 
general information. 
 
Since the dispersion-theoretical evaluation is restricted to the 
contribution of (correlated) $\pi\pi$ and $K\anti{K}$ exchange to the
baryon-baryon amplitudes only those singularities are taken into 
account which are generated by $\pi\pi$ and $K\anti{K}$ intermediate 
states, namely the discontinuities due to the
$\pi\pi$ and $K\anti{K}$ unitarity cut (the so-called right-hand cut). 
The left-hand cuts, which are due to unitarity constraints
for the $u$-channel reaction, can be neglected in the baryon-baryon
channels considered here, since they start at large, negative
$t$-values (from which they extend to $-\infty$) and are therefore far
away from the physical region relevant for low-energy $s$-channel
processes.
 
The $B\anti{B'}\to\alpha$ amplitudes, which enter
in Eq.~(\ref{unitarity}) are derived from a microscopic model which is
based on the hadron-exchange picture, cf. Sect. II. 
Of course, this model has a limited range of validity: for energies
far beyond $t'_{max}\approx 100\,m_\pi^2$ it cannot provide reliable
results.
The dispersion integral for the invariant amplitudes extending in
principle along the whole $\pi\pi$ right-hand cut has therefore to be
limited to an upper bound, $t'_{max}$, which has been put to 
$t'_{max}$ = 120~$m_\pi^2$ in Ref. \cite{REUBER}. 
  
The spectral function (\ref{spectral}) for the ($0^+$) $\sigma$-channel 
has only one component but the one for the ($1^-$) $\rho$-channel
consists of four linearly independent components, which reflects
the more complicated spin structure of this channel. 

\begin{figure}[hb]
\vskip 4.3cm 
\includegraphics{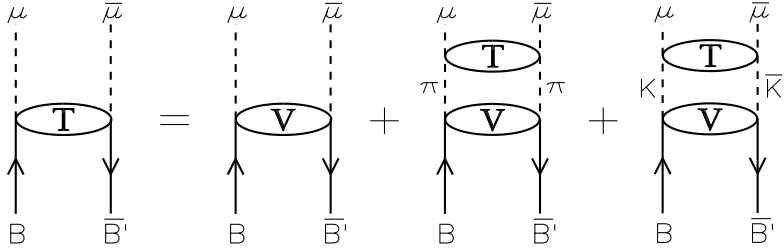} 
\caption{The dynamical model for the
        $B\bar B \rightarrow \mu \bar \mu$
        amplitude ($\mu \bar \mu$ = $\pi\pi$, $K\bar K$).}
\label{fig5}
\end{figure}

\begin{figure}[hb]
\vskip 4.3cm 
\includegraphics{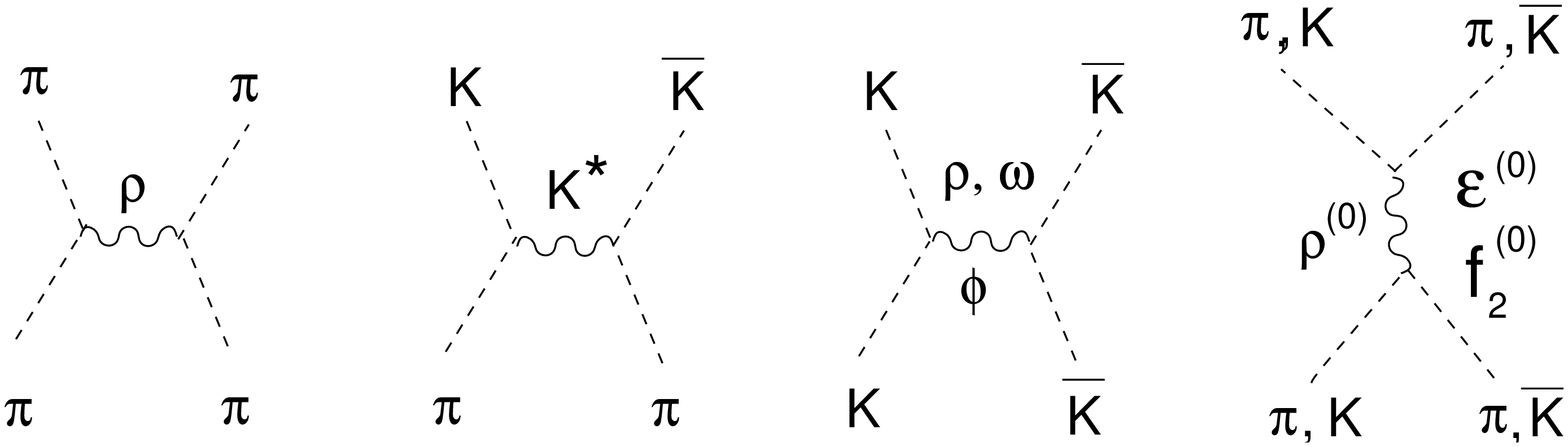}
\caption{The contributions to the potential of the coupled
channel $\pi\pi - K\bar K$ model of Ref. \protect\cite{Schutz}.
}
\label{pipi}
\end{figure}

\begin{figure}[t]
\vglue 1cm
\includegraphics{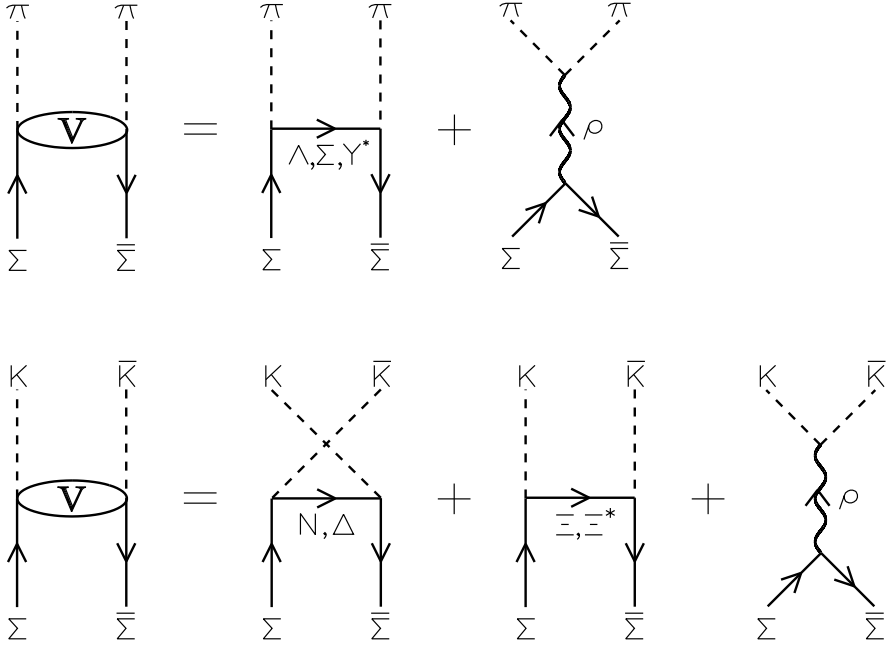}
\vskip 12cm
\caption{The contributions to the Born amplitude for the
transitions $\Sigma \anti{\Sigma} \to \pi\pi, K\bar K$.
}
\label{Born}
\end{figure}

Finally, 
we should note that the helicity amplitudes obtained according to
Fig.~\ref{fig5} still generate the uncorrelated (first diagram on
the r.h.s. of Fig.~\ref{fig5}), as well as the correlated pieces
(second and third diagrams).
Thus, in order to obtain the contribution of the truely correlated
$\pi\pi$ and $K\anti{K}$ exchange one must eliminate the former from
the spectral function.
This is done by calculating the spectral function generated by
the Born term and subtracting it from the total spectral function:
\begin{equation}
\rho^{(0^+,1^-)} \longrightarrow \rho^{(0^+,1^-)} - 
\rho^{(0^+,1^-)}_{\rm Born} .
\end{equation}
In practice this means that, e.g., for the full Bonn $NN$ model 
contributions involving spin-1/2 as well as spin-3/2 baryons 
have to be subtracted since corresponding contributions are already
treated explicitly in the s-channel in this model, namely via box 
diagrams with intermediate $\Delta$-states as shown in 
Fig.~\ref{fig1}(a). On the other hand only
uncorrelated contributions involving spin-1/2 baryons are to be 
subtracted from the discontinuities of the invariant baryon-baryon
amplitudes in order to avoid double counting if a simple 
OBE-model is used in the $s$-channel. This is the relevant procedure
for the $YN$ model that will be presented in the next section. 

Note that the spectral functions characterize both the strength
and range of the interaction.
Clearly, for sharp mass exchanges the spectral function becomes
a $\delta$-function at the appropriate mass. 

For convenience the authors of Ref. \cite{REUBER} have presented 
their results in terms of effective coupling strengths, by 
parameterizing the correlated processes
by (sharp mass) $\sigma$ and $\rho$ exchanges.
The interaction potential resulting from the exchange of a
$\sigma$ meson with mass $m_\sigma$ between two $J^P=1/2^+$
baryons $A$ and $B$ has the structure:
\begin{equation}
V^{\sigma}_{A,B \to A,B}(t) \ = \ g_{AA\sigma} g_{BB\sigma} 
{F^2_\sigma (t) \over t - m^2_\sigma} , 
\label{formd}
\end{equation}
where a form factor $F_\sigma(t)$ is applied at each vertex,
taking into account the fact that the exchanged $\sigma$ meson is
not on its mass shell. 
This form factor is parameterized in the conventional monopole form, 
\begin{equation}
F_\sigma (t) = {\Lambda ^2_\sigma - m^2_\sigma \over 
\Lambda ^2_\sigma - t} \ , 
\label{form}
\end{equation}
with a cutoff mass $\Lambda_\sigma$ assumed to be the same
for both vertices.
The correlated potential as given in Eq.~(\ref{dispersion}) can now be
parameterized in terms of $t$-dependent strength functions
$G_{B_1',B_2' \to B_1,B_2}(t)$, so that for the $\sigma$ case:
\begin{equation}
V^{(0^+)}_{A,B \to A,B}(t) = 
G^{\sigma}_{AB \to AB}(t) F^2_\sigma(t) {1 \over t - m^2_\sigma}. 
\label{sigma}
\end{equation}
The effective coupling constants are then defined as:
\begin{equation}
g_{AA\sigma}g_{BB\sigma} \quad\longrightarrow \quad G_{AB\to
AB}^\sigma (t)= {(t-m_\sigma^2)\over\pi F^2_\sigma(t)}
\int_{4m_\pi^2}^{\infty} {\rho^{(0^+)}_{AB \to AB}(t') \over t'-t} dt' .
\label{effccsig}
\end{equation}

Similar relations can be also derived for the correlated exchange
in the isovector-vector channel \cite{REUBER}, which in this case
will involve vector as well as tensor coupling pieces.

It should be stressed that, so far, this parameterization does not involve 
any approximations as long as the full $t$-dependence of the effective
coupling strengths is taken into account.
The parameters of the $\sigma$ and $\rho$ exchange have been chosen to have
the same values in all particle channels.
The masses $m_\sigma$ and $m_\rho$ of the exchanged particles have
been set to the values used in the Bonn-J\"ulich models of the
$NN$~\cite{MHE} and $YN$~\cite{Holz} interactions,\ 
$m_\sigma=550$ MeV, $m_\rho=770$ MeV.
The cutoff masses $\Lambda_{\sigma}$ and $\Lambda_{\rho}$ have been
chosen so that the coupling strengths in the $S=0, -1$ baryon-baryon
channels vary only weakly with $t$.
The resulting values ($\Lambda_\sigma=2.8$ GeV, $\Lambda_\rho=2.5$ GeV)
are quite large compared to the values of the phenomenological
parameterizations used in Refs.~\cite{Holz,MHE}, and thus represent
rather hard form factors.

Note that in the OBE framework the contribution of a genuine
(SU(3)) $\sigma$ meson to the three reactions $NN\rightarrow NN$, 
$YN\rightarrow YN$, $YY \rightarrow YY$ is determined by two parameters 
(coupling constants), namely $g_{NN\sigma}$
and $g_{YY\sigma}$, whereas the correlated exchange is 
characterized by three independent strength functions 
($G_{NN\to NN}$, $G_{YN\to YN}$, $G_{YY\to YY}$) so that vertex
coupling constants cannot be determined uniquely. This implies directly
that the strength parameters cannot fulfill SU(3) relations.

In the physical region the strength of the contributions is to a large
extent governed by the value of $G$ at $t=0$. 
Those values for the various channels were tabulated in Ref. \cite{REUBER}
(cf. Tables 5-7) for the case of the full model calculation 
and also 
when uncorrelated contributions involving spin-1/2 baryons only
are subtracted from the spectral function of the invariant baryon-baryon
amplitudes. The latter are the proper values to be used for constructing
a $YN$ model based on simple OBE-exchange diagrams. 

\begin{table}[h]
\begin{tabular}{|rcccccccc|}
\hline
&\multicolumn{8}{c|}{$G_{YN\to Y'N}/4\pi$} \\
\hline
\hline
&\multicolumn{4}{c}{$\Lambda N$}&\multicolumn{4}{c|}{$\Sigma N$}\\
\hline
\mbox{$\sigma$ channel} &\multicolumn{4}{c}{3.52}&\multicolumn{4}{c|}
        {2.92}\\
\hline
\hline
&\multicolumn{4}{c}{$\Sigma N$}&\multicolumn{4}{c|}{$\Lambda N \to \Sigma N$}\\
\hline
&VV & VT & TV & TT & VV & VT & TV & TT \\ 
\hline
\mbox{$\rho$ channel} & \ 1.26 \ & \ 1.24 \ & \ 8.09 \ & \ 8.07 \ & 
  \ $-$0.43 \ & \ 3.72 & \ $-$1.32 \ & \ 21.00 \\
\hline
\end{tabular}
\caption{Effective $\sigma$ and $\rho$ coupling strengths $G_{YN\to Y'N}(t=0)$
        for correlated $\pi\pi$ and $K\anti{K}$ exchange in the various
        nucleon-hyperon channels. $VV$, $VT$, etc. stand for the vector-vector,
        vector-tensor, etc. combinations of the $\rho$ coupling, cf. 
        Ref.~\cite{REUBER}.}
\label{coup0} 
\end{table}

In principle, 
the average size of the effective coupling strengths is only an
approximate measure of the strength of correlated $\pi\pi$ and
$K\anti{K}$ exchange in the various particle channels.
The precise energy dependence of the correlated exchange as well
as its relative strength in the different partial waves of the
$s$-channel reaction is determined by the spectrum of exchanged
invariant masses, or spectral functions, leading to a different
$t$-dependence of the effective coupling strengths.
This was demonstrated in Ref. \cite{Melni1} where the on-shell
$NN$, $\Lambda N$ and $\Sigma N$ potentials in spin-singlet states 
with angular momentum $L=0, 2$ and 4, generated 
directly by the scalar-isoscalar part of the correlated $\pi\pi$ and 
$K\anti{K}$ exchange, were compared to the corresponding 
results based on a $\sigma$ exchange with sharp mass. 
It could be seen that the correlated $2\pi$ exchange is significantly stronger 
in high partial waves because
the $\sigma$ exchange, which corresponds to a spectral function proportional
to $\delta(t'-m^2_\sigma)$, does not contain the long-range part of the
correlated processes. Thus, parameterizing the results derived from
the microscopic model by $\sigma$ exchange with a sharp mass, but using 
the effective coupling strength $G^\sigma_{NN\to NN}$ at $t=0$ one can 
obtain rough agreement with the exact result in the $S$ waves, say,
but usually underestimates the magnitude considerably in the high partial 
waves. Obviously the replacement of correlated $\pi\pi$ and $K\anti{K}$
exchanges by an exchange of a sharp mass $\sigma$ meson with a
$t$-independent coupling cannot provide a simultaneous description
of both low and high partial waves.

These features are important for investigations of the $NN$ systems where
the phase shifts are known quantitatively even for rather high partial waves.
In this case the results of the correlated exchange should be used
directly \cite{Kim}. However, for the $\Lambda N$ and $\Sigma N$ systems
only scattering observables are available, and those (total and differential cross 
sections) are primarily sensitive to $S$- and $P$-wave contributions. 
Thus, here it is reasonable to simplify the calculation and use only an
effective parametrization of the results derived from
the microscopic model in terms of a $\sigma$ and $\rho$ exchange with a sharp 
mass. Specifically, combining Eqs. (\ref{formd}) and (\ref{sigma}) we use the 
expression 
\begin{equation}
V^{(0^+)}_{A,B \to A,B}(t) = 
G^{\sigma}_{AB \to AB} \tilde F_{\sigma}^2 (t) {1 \over t - m^2_\sigma}. 
\label{corrpot}
\end{equation}
with 
\begin{equation}
\tilde F_{\sigma} (t) = {\Lambda ^2_{\sigma} \over 
\Lambda ^2_{\sigma}- t} \  
\label{form1}
\end{equation}
and a similar one for the $\rho$ exchange contribution. 
The effective coupling strength $G^\sigma_{YN\to Y'N}$ (and 
$^{ij}G^\rho_{YN\to Y'N}$) is deduced via Eq. (\ref{effccsig}) 
(and via a similar one for the $\rho$ channel, cf. Ref. \cite{REUBER}) for the
form factor (\ref{form1}) and adjusted to the value at $t=0$. 
The different prescription for the vertex form factor as compared to Ref. \cite{REUBER},
i.e to Eq. (\ref{form}), is adopted here because it guarantees that the 
on-shell behaviour of the potential (which is fully determined by the
dispersion integral) is not modified strongly as long as the energy is not too high. 
At the same time smaller cutoff masses as those mentioned above (and employed in 
\cite{REUBER}) can be used to ensure sufficient convergence when the potential 
(\ref{corrpot}) is iterated in the scattering equation. The concrete values
used for the cutoff masses are $\Lambda_\sigma$ = 2.5 (1.6) GeV for the
$\Lambda N$ ($\Sigma N$) channels and $\Lambda_\rho$ = 1.25 (1.8) GeV for the
$\Lambda N \to \Sigma N$ transition ($\Sigma N$ channel). 

The effective coupling strengths employed in our new $YN$ model are compiled in 
Table \ref{coup0}. Though these values differ slightly from those given
in Tables 5-7 of Ref.~\cite{REUBER}, due to the different choice
of the form factor, we would like to emphasize that the strengths
of the interactions at $t=0$ are the same in both cases and
coincide with the one derived from the microscopic model of
$\pi\pi$ and $K\bar K$ correlations. 

\begin{figure}[hb]
\vskip 5cm 
\includegraphics{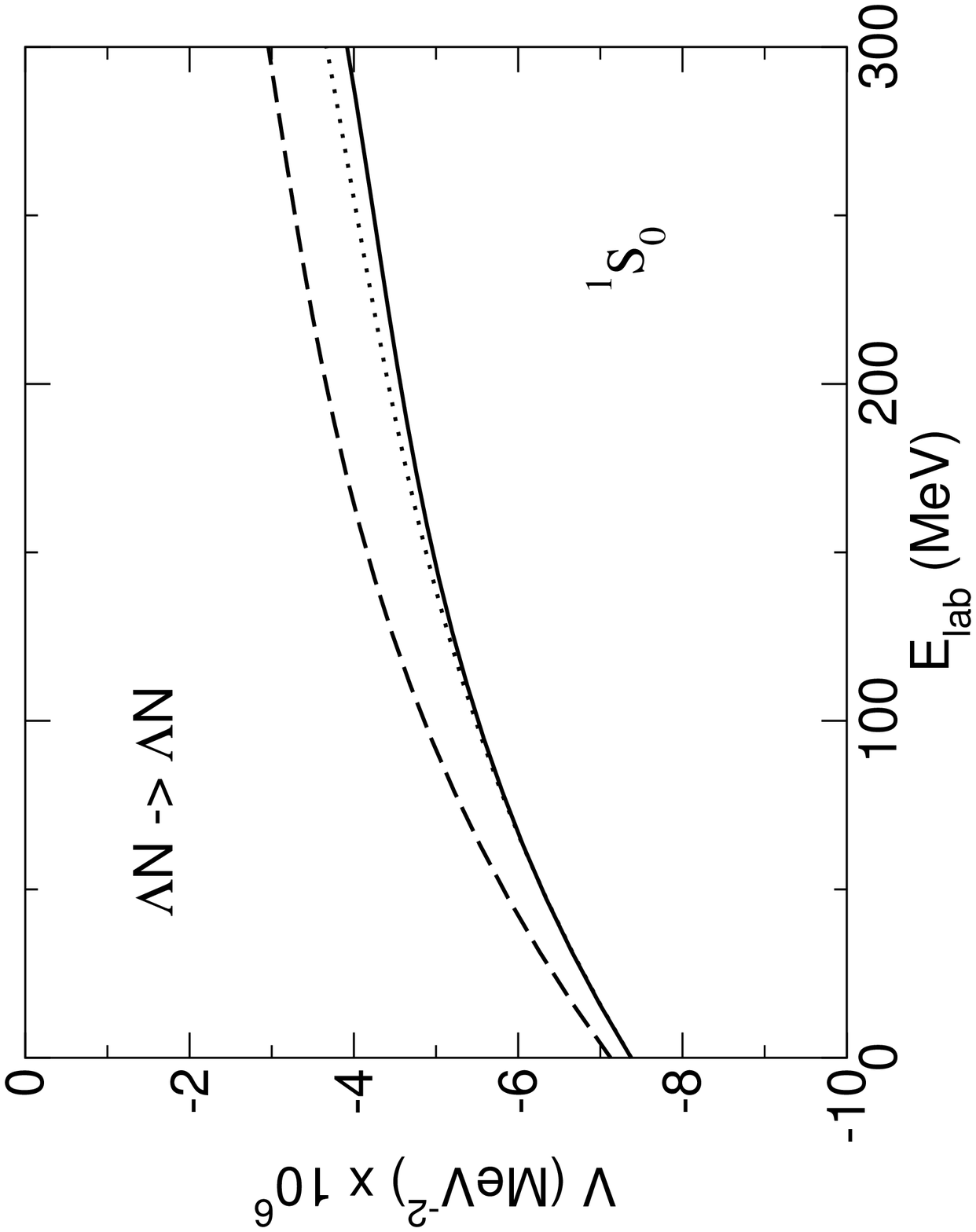} 
\includegraphics{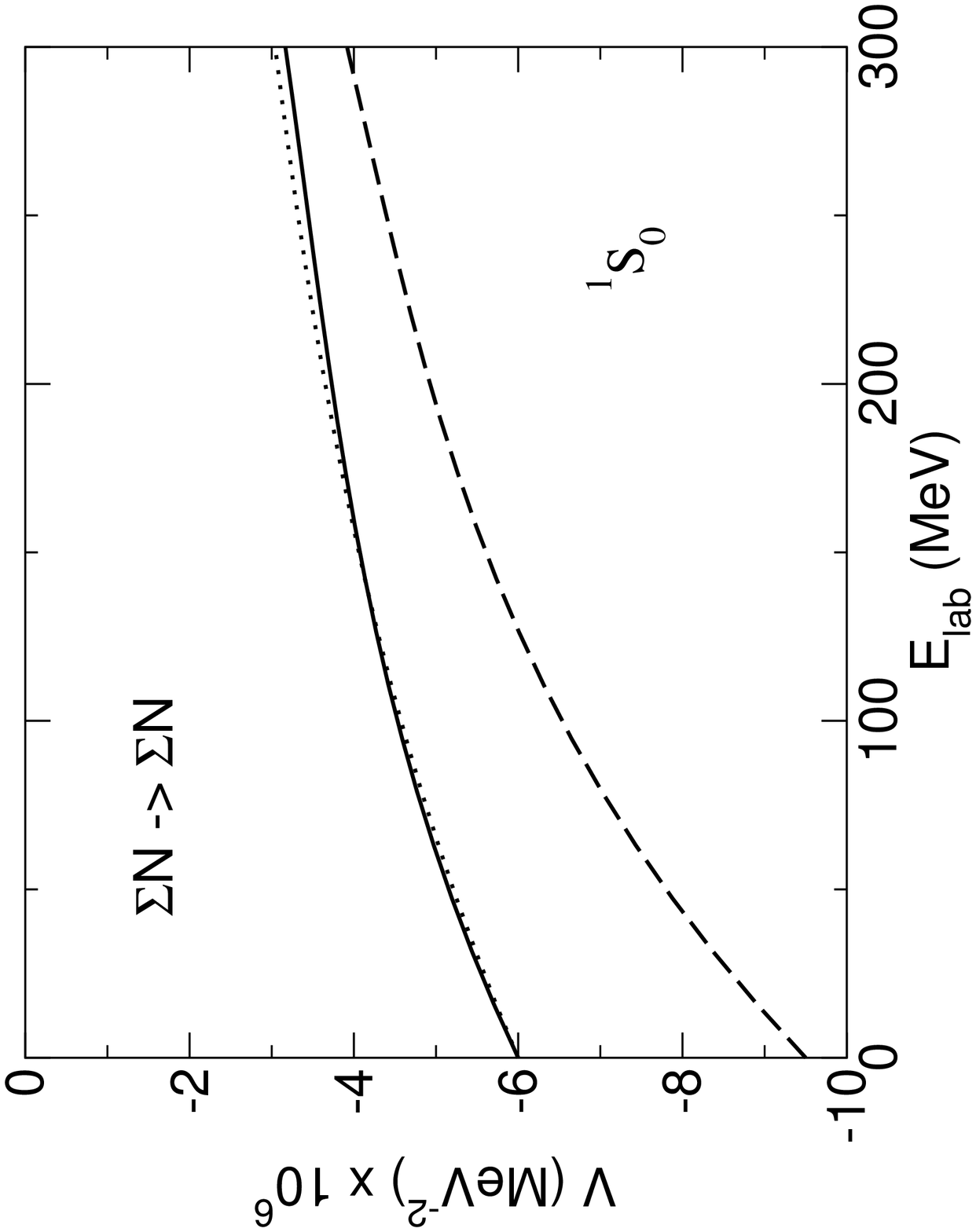} 
\caption{{The $\sigma$-like part of the $\Lambda N$ and $\Sigma N$ 
on-shell potentials in the $^1S_0$ partial wave. The solid lines are 
derived from our microscopic model of correlated $\pi\pi$ and $K\bar K$ 
exchange. The dotted lines are obtained if the dispersion-theoretical
result is parameterized by $\sigma$ exchange, while the dashed lines 
correspond to the $\sigma$ exchange used in the J\"ulich $YN$ potential $A$
\protect\cite{Holz}. 
}} 
\label{fig:5_6_4}
\end{figure}

In order to demonstrate that, we show in Fig.~\ref{fig:5_6_4} the 
corresponding on-shell potential matrix elements 
for the $^1S_0$ partial wave of the $\Lambda N$ and $\Sigma N$ channels.
One can see that in case of the $\Lambda N$ system the result generated by 
the scalar-isoscalar part of correlated $\pi\pi$ and $K\anti{K}$ exchange 
is similar to the one of the $\sigma$ exchange used in the 
J\"ulich $YN$ model~A. In fact, correlated $\pi\pi$ exchange is 
marginally stronger. It is also obvious that the parameterization of
the interaction generated by correlated $\pi\pi$ exchange by 
an effective $\sigma$ exchange, c.f. the dotted line, works rather well. 
{}From the corresponding results for the on-shell $\Sigma N$ potential 
one can see that here the $\sigma$ exchange used in the J\"ulich $YN$
model A is clearly much stronger than what one obtains from
the correlated $\pi\pi$ and $K\anti{K}$ exchange. Once again the
parameterization by an effective $\sigma$ exchange provides an
excellent representation of the interaction strength. 
 
\begin{table}[ht]
\caption{Vertex coupling constants used in the new $YN$ model that are 
constrained by SU(6) symmetry and corresponding cutoff masses. The
assumed SU(6) symmetry fixes the $F/(F+D)$ ratios to 
$\alpha_{ps}$=2/5, $\alpha_{v}^e$=1, $\alpha_{v}^m$=2/5 \cite{Reu}. 
}
\label{coup1}
\begin{center}
\begin{tabular}{|c|ccc|}
\hline
 Vertex & $g_{BB'm}/\sqrt{4\pi}$ & $f_{BB'm}/\sqrt{4\pi}$ & $\Lambda_{BB'm}$ (GeV) \\
\hline
 $NN\pi$  & 3.795 &  &  1.3 \\
 $\Lambda\Sigma\pi$  & 2.629 &  &  1.3 \\
 $\Sigma\Sigma\pi$  & 3.036 &  &  1.3 \\
 & & & \\
 $N\Lambda K$ & $-$3.944 &  &  1.2 \\
 $N\Sigma K$  & 0.759 &  &  1.2 \\
 & & & \\
 $NN\omega$  & 3.317 &  &  1.7 \\
 $\Lambda\Lambda\omega$  & 2.211 & $-$2.796  &  1.4 \\
 $\Sigma\Sigma\omega$  & 2.211 & 2.796  &  1.7 \\
 & & & \\
 $N\Lambda K^*$ & $-$1.588 & $-$5.175 &  1.2 \\
 $N\Sigma K^*$  & $-$0.917 & 2.219 &  1.4 \\
\hline
\end{tabular} \end{center}
\end{table}

%%%%%%%%%%%%%%%%%%%%%%%%%%%%%%%%%%%%%%%%%%%%%%%%%%%%%%%%%%%%%%%%%%%%%%%%%
\section{Results and discussion}

\subsection{Coupling constants}

In the present $YN$ model we take into account exchange diagrams 
involving the well-established lowest lying pseudoscalar and vector
meson SU(3) octets. Following the philosophy of the original J\"ulich
$YN$ potential \cite{Holz} the coupling constants in the pseudoscalar
sector are fixed by strict SU(6) symmetry. In any case, this is 
also required for being consistent with the model of correlated $\pi\pi$ 
and $K\bar K$ exchange. The cutoff masses of the
form factors belonging to the $NN$ vertices are taken over from the
full Bonn $NN$ potential. The cutoff masses at the strange vertices
are considered as open parameters though, in practice, their values
are kept as close as possible to those found for the $NN$ vertices,
cf. Table \ref{coup1}. Note that like in \cite{Holz} and in line with
the arguments brought forth in Ref. \cite{Reu} we neglect again 
the contribution from $\eta$ meson exchange. Anyhow, in the full
Bonn $NN$ model the $\eta NN$ coupling constant was set to zero.
In addition phenomenological analyses \cite{Grein} and also 
microscopic calculations, like those based on the topological chiral
soliton model (extended Skyrme model with vector mesons)
\cite{etaNN}, indicate that this coupling constant should be small. Thus, 
the $\eta$ contribution would be completely unimportant anyway, 
given the pseudoscalar nature of its coupling. For the same reason
the $\eta'$ contribution is likewise not considered. 

In the vector meson sector we depart from the strategy of the original 
J\"ulich $YN$ potential. As already mentioned above, first and most 
importantly the contribution of the $\rho$ meson is no longer seen
as resulting from the exchange of a genuine particle that belongs to 
the SU(3) vector meson octet but is identified with the strength 
generated by a microscopic model of correlated $\pi\pi$ and $K\bar{K}$ 
in the vector-isovector channel. The effective coupling constants for 
$\rho$ exchange in 
the various $YN$ and $YY$ channels have been extracted and thoroughly
analysed in Ref. \cite{REUBER}. Thereby it was found that the result from
correlated exchange deviates significantly from those implied by SU(3)
symmetry -- even though SU(3) symmetry was imposed for the bare $\rho NN$
and $\rho YY$ couplings, cf. \cite{REUBER}. 
In view of this it is questionable whether one should invoke SU(3) symmetry 
for fixing the other coupling strengths of the vector-meson octet, i.e.
those of the $K^*$ meson and of the coupling of the $\omega$ meson to the hyperons. 
But in absence of any better alternative we still follow this prescription 
for the present model. As reference values we take here the $NN\rho$ coupling
constants of the full Bonn $NN$ potential \cite{MHE}, which were already used
for the old $YN$ model \cite{Holz,Reu}.
However, as far as the $YY\omega$ coupling constants are concerned now 
we take into account the insight gained in Ref. \cite{Janssen1} that 
the $\omega$ exchange in the full Bonn $NN$ potential represents not only the 
genuine SU(3) $\omega$ but is also an effective parametrization of additional 
short-range contributions from correlated $\pi-\rho$ exchange, say, that are not 
included explicitly in that model. Therefore, in the Bonn $NN$ model 
the required $NN\omega$ coupling constant is indeed much larger than what 
follows from the SU(3) relations and this large coupling constant formed also 
the basis for fixing the $YY\omega$ coupling constants of the old J\"ulich
$YN$ model \cite{Holz,Reu}, cf. the discussion in Sect. 2.2 of Ref. \cite{Reu}.
In the present model we adopt the smaller value
found in Ref. \cite{Janssen1} which is very close to the SU(3) value.
This is in line with results obtained from a dispersion-theoretical analysis
of the nucleon electromagnetic form factors - the inclusion of the $\pi-\rho$
continuum sizeably reduces the  $\omega NN$ coupling, compare the values
found in \cite{MMD}  with the ones in \cite{MMSvO}.
Assuming furthermore that the $\rho$ meson couples universally to the
isospin current -- which fixes the $F/(F+D)$ ratio $\alpha^e_V$ to 1 --
and ideal mixing for the $\phi$ and $\omega$ mesons then yields the
following relation for the $\omega$ coupling constants:
\begin{eqnarray}
g_{\Lambda\Lambda\omega}=g_{\Sigma\Sigma\omega} = {2\over 3} g_{NN\omega}, \ \ \ 
f_{\Lambda\Lambda\omega}={5\over 6} f_{NN\omega} 
-{1\over 2} f_{NN\rho}, \ \ \ 
f_{\Sigma\Sigma\omega}={1\over 2} f_{NN\omega} +{1\over 2} f_{NN\rho} 
\end{eqnarray}
For $f_{NN\omega}$ and f$ _{NN\rho}$ we take over the values of 
the full Bonn $NN$ potential. Since $f_{NN\omega}$=0 \cite{MHE}
it follows that $f_{\Lambda\Lambda\omega} =- f_{\Sigma\Sigma\omega}$.
 
The short-range contributions from correlated $\pi-\rho$ exchange were
parametrized by an effective $\omega '$ exchange in Ref. \cite{Janssen1} 
with a mass of $m_{\omega '}$ = 1120 MeV. We follow here the same strategy
but treat the coupling constants of the $\omega '$ to the strange baryons
as free parameters to be determined in a fit to the $YN$ data.  

Like the $\rho$ also the contribution of the $\sigma$ meson is 
computed from a microscopic model of correlated $\pi\pi$ and $K\bar{K}$ 
--  now from the scalar-isoscalar channel. The effective coupling constants 
for $\sigma$ exchange in the various $YN$ channels have been discussed
in the previous section. 

\begin{table}[ht]
\caption{Parameters (effective coupling strengths $G$, cutoff masses $\Lambda$) 
used in the new $YN$ model for the effective $\omega '$, $a_0$ and $\kappa$ exchanges. 
In case of $\omega '$ only the vector-vector component is considered .
Cutoff masses in parentheses indicate that here a product of form factors of 
monopole type (\ref{form}) is utilized instead of the standard dipole form, 
cf. Eq. (\ref{formd}).
Numbers in square brackets denote corresponding values of the model where
$\kappa$ exchange is replaced by a contact term, cf. text, when different. 
}
\label{coup2}
\begin{center}
\begin{tabular}{|ccccc|}
\hline
 channel & exchange & mass (GeV) & $G_{BB'\to BB'}/(4\pi)$ &  $\Lambda_{BB'm}$ (GeV) \\
\hline
$\Lambda N$ & $\omega '$ & 1.12 & 5.0 &  1.65 \\
 & $\kappa$ & 1.0 & 6.0 [1.8] &  1.45 [1.5] \\
 & & & & \\
$\Lambda N \to \Sigma N$ 
 & $a_0$ & 0.983 & 2.0 & 2.0 (1.8) [2.0 (2.1)] \\
 & $\kappa$ & 1.0 & 6.0 [1.9] & 1.45 (1.65) [1.5] \\
 & & & & \\
$\Sigma N$ & $\omega '$ & 1.12 & 10.75 & 1.35 \\
 & $a_0$ & 0.983 & 5.63 & 2.0 (1.45) \\
 & $\kappa$ & 1.0 & 6.0 [2.4] & 1.65 [1.5] \\
\hline
\end{tabular} \end{center}
\end{table}

Besides replacing the conventional $\sigma$ and $\rho$ exchanges by
correlated $\pi\pi$ and $K\bar{K}$ exchange, there are in addition
some other new ingredients in the present $YN$ model.
First of all, we now take into account contributions from $a_0(980)$
exchange.
The $a_0$ meson is present in the original Bonn $NN$ potential
\cite{MHE}, and for consistency should also be included in the $YN$
model.
Secondly, we consider the exchange of a strange scalar meson, the
$\kappa$, with mass $\sim 1000$~MeV.
Let us emphasize, however, that like in case of the $\sigma$ meson 
these particles are not viewed as being members of a scalar meson SU(3) 
multiplet, but rather as representations
of strong meson-meson correlations in the scalar--isovector
($\pi\eta$--$K\bar K$) \cite{Janssen} and scalar--isospin-1/2 ($\pi K$)
channels \cite{Lohse}, respectively.
In principle, their contributions can also be evaluated along the lines
of Ref.~\cite{REUBER}, however, for simplicity in the present model they
are effectively parameterized by one-boson-exchange diagrams with the
appropriate quantum numbers assuming the coupling constants to be free
parameters. The parameters specifying those ingredients are summarized
in Table \ref{coup2}. 

\begin{figure}[t]
\vglue 1cm
\includegraphics{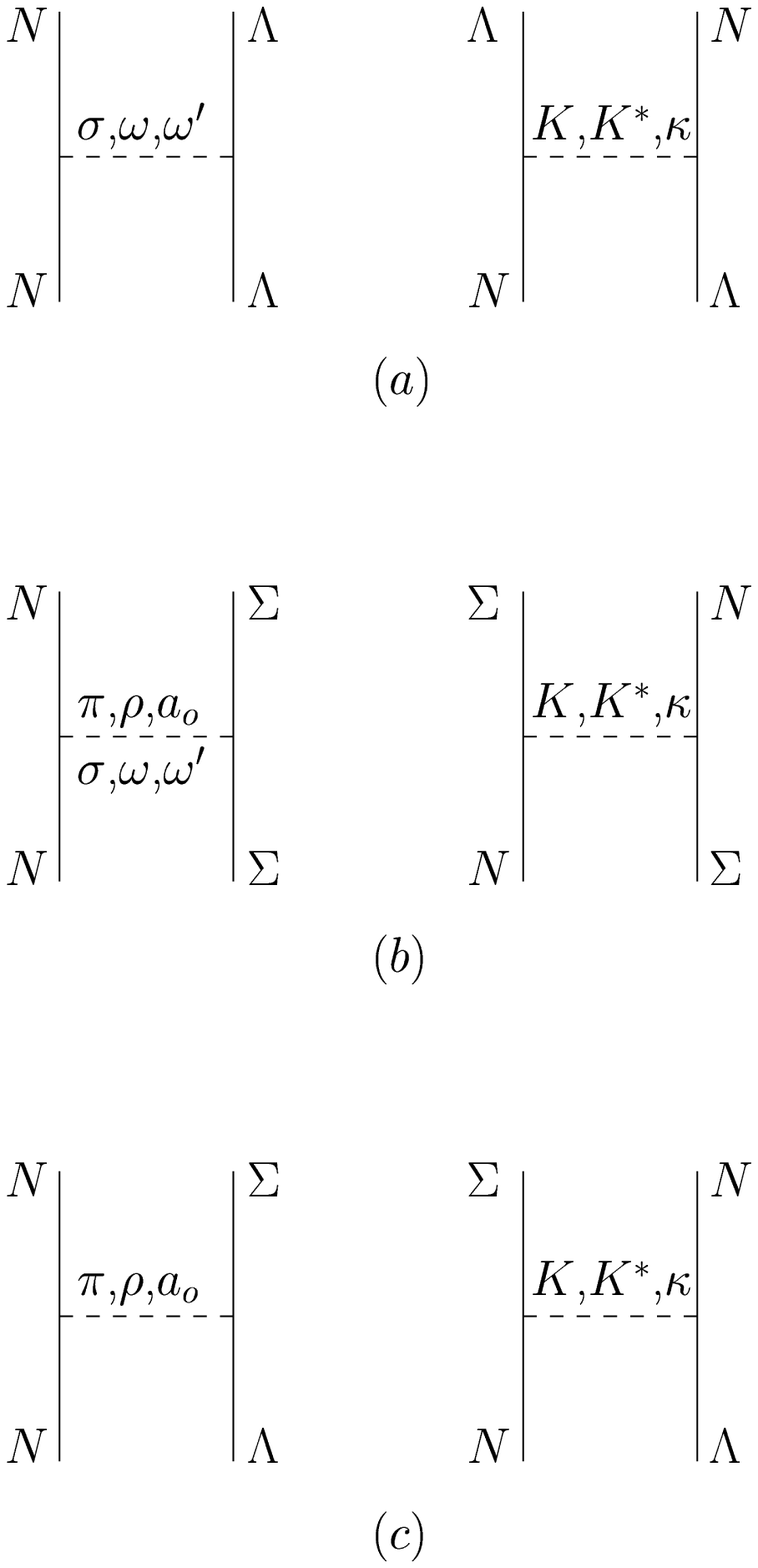} 
\vskip 18cm
\caption{Contributions to the interaction in the
(a) $\Lambda N$ and (b) $\Sigma N$ channels and in the 
(c) $\Lambda N \to \Sigma N$ transition. Note that only 
$\pi$, $K$, $\omega$, and $K^*$ exchange are considered as being 
due to genuine SU(3) mesons. The other contributions are
either fixed from correlated $\pi\pi$ and $K\bar K$ exchange
($\sigma$, $\rho$) or are viewed as an effective parametrization
of meson-meson correlations ($a_0$, $\kappa$, $\omega$')
in the corresponding spin-isospin channels. 
}
\label{figyn}
\end{figure}

Thus we have the following scenario: The long- and intermediate-range part 
of our new $YN$ interaction model is completely 
determined by SU(6) constraints (for the pseudoscalar and to  
some extent also for the vector mesons) and by correlated 
$\pi\pi$ and $K\bar K$ exchange. The short-range part is viewed as
being also due to correlated meson-meson exchanges but in practice is 
parametrized phenomelogically in terms of one-boson-exchange 
contributions in specific spin-isospin channels. In particular, 
no SU(3) relations are imposed on the short-range part. This 
assumption is based on our observation that the contributions in the 
$\rho$ exchange channel as they result from  
correlated $\pi\pi$ and $K\bar{K}$ no longer fulfill
SU(3) relations, but it also acknowledges 
the fact that at present there is no general agreement about who are
the actual members of the lowest-lying scalar meson SU(3) multiplet.
A graphical representation of all meson-exchange contributions that are
included in the new $YN$ model is given in Fig. \ref{figyn}. 

In recent investigations of the $NN$ interaction within the framework
of chiral perturbation theory \cite{Epelbaum}
only pionic degrees of freedom are taken into account and all short-range 
physics is parametrized by contact terms. This is certainly also an 
option that one should explore for the $YN$ system \cite{Korpa} in the future \cite{Henk}.
As a first step we consider here an alternative model where the 
contributions of the $\kappa$(1000) meson - whose mass and even existence is
still under dispute \cite{Kappa} -- are substituted by a contact term.   
In practice this means that we replace the product of the $\kappa$ coupling
constants and propagator, $G_{BB'\to BB'}/(m_\kappa^2-t)$,
by $G_{BB'\to BB'}/m_\kappa^2$ and readjust only the parameters of
related to the $\kappa$ exchange (with one exception). Those parameters 
can be found also in Table \ref{coup2}, in square brackets, for those 
cases where they differ from the values of our regular model.  
Results corresponding to the model with the contact term will also be 
presented in the next section. 

In the fitting procedure we only take into account data on total 
cross sections (and energies near the corresponding thresholds)
for the channels
$\Lambda p$ \cite{Alex,Sechi,Kadyk}, $\Sigma^-p$ \cite{Eisele},
$\Sigma^-p \to \Lambda n$ \cite{Engel}, $\Sigma^-p \to \Sigma^0n$ \cite{Engel},
and $\Sigma^+p$ \cite{Eisele}. 
Differential cross sections but also total cross sections at higher energies 
\cite{Stephen,Kondo,Ahn} are therefore genuine predictions of our model. 
As already mentioned above, the free parameters in our model consist of
the cut-off masses at the strange vertices and the coupling constants 
of the $a_0$(980), $\kappa$(1000) and $\omega '$(1120) mesons. 
When adjusting those parameters to the empirical data it turned
out that the results are not very sensitive to the cut-off masses in
the pseudo-scalar sector and we fixed them to be close to the cutoff mass
used at the $\pi NN$ vertex. There is also only a weak sensitivity to the
cut-off masses used for the correlated $\pi\pi$-$K\bar K$ contributions
in the $\sigma$ and $\rho$ channels. This is due to the chosen analytic
form of the form factors that practically does not change the strength of 
the corresponding potentials as they result from the microscopic model 
-- which is of course intended, cf. the discussion in Sect. III. 
Besides the cut-off masses of the vector mesons we found that also 
the parameters of the $a_0$(980) and $\kappa$(1000) mesons, 
viewed here as effective parametrization of correlated $\pi\eta$ and
$\pi K$ exchange, have a sizeable influence. In fact, without the
contributions of the latter two mesons we would not have been able
to achieve a satisfactory description of the data. Note that those
two exchanges were not considered in the original J\"ulich model \cite{Holz}. 
We should say that values of the coupling strengths and cut-off masses for
those scalar mesons are strongly correlated and cannot be fixed independently
from a fit to the data. Thus, one should not attribute any physical
significance to the actual values of the coupling strengths or cut-off 
masses that we found individually. 

Finally we want to mention that the fit to the available $YN$ data did not
constrain the relative magnitude of the $^1S_0$ and $^3S_1$ partial waves
in the $\Lambda N$ system. Thus, as a further constraint, we required the
$^1S_0$ scattering length to be larger than the one for $^3S_1$ -- as
it seems to be necessary if one wants to achieve a bound 
hypertriton \cite{Ueda}. A first application of the new $YN$ model in 
three-body calculations confirmed that it yields indeed a bound 
hypertriton state \cite{Nogga}.

\subsection{The scattering equation} 

The original $YN$ model of the J\"ulich group was derived within the 
framework of time ordered perturbation theory (TOPT) \cite{Holz}. In 
this approach retardation effects from the meson-exchange diagrams
are retained (and those of baryon-exchange as well) and as 
a consequence the interaction depends explicitly on the starting
energy. This is not convenient if one wants to apply the $YN$ model 
in conventional few-body 
\cite{Miya1,Miya2,Miya4,Akaishi,Nogga1,Nemura,Fujiwara3N,Hiyama}
or many-body \cite{Ramos,Tzeng,Fujii,Lenske} investigations. 
Thus, in Ref. \cite{Reu} the
J\"ulich group presented energy-independent versions of their $YN$ 
model where the energy dependence was removed in such a way that
basically all other characteristics of the original model could be
kept. The detailed comparison of the TOPT model and its 
energy-independent counterpart performed in Ref. \cite{Reu} made
clear that this goal was indeed achieved. 

Since we are also interested to facilitate an application of our 
new $YN$ model in future few- and many-body investigations we will 
likewise present here an energy-independent interaction. This 
implies that we do not use the (relativistic)
TOPT scattering equation of Ref. \cite{Holz} but instead solve
the nonrelativistic (coupled-channel) Lippmann-Schwinger equation
\begin{equation}
T_{i,j} = V_{i,j} + \sum_k V_{i,k} G_k T_{k,j} 
\label{tmat}
\end{equation}
to obtain the scattering amplitude $T_{i,j}$. Here the indices
($i,j,k$) stand for the $\Lambda N$ and $\Sigma N$ channels
and the nonrelativistic Green's function $G_k$ is given 
by 
\begin{equation}
G_k = \left[ {q_k^2-{\bf q}'^2 \over {2\mu_k}} + i\varepsilon \right]^{-1} \ , 
\label{Green}
\end{equation}
where $\mu_k = M_YM_N/(M_Y+M_N)$ is the reduced mass and ${\bf q}'$ the
c.m. momentum in the intermediate $YN$ channel. $q_k = q_k(z)$ denotes the
on-shell momentum in the intermediate $YN$ state defined by 
$z = \sqrt{M_Y^2 + q_k^2} + \sqrt{M_N^2 + q_k^2}$. The latter equation 
guarantees that the $\Sigma N$ channel opens exactly at the physical
threshold.  Note that $q_{\Sigma N}$ is imaginary for starting energies 
below the $\Sigma N$ threshold ($z < M_\Sigma + M_N$). 
Explicit expressions for the potential matrix elements $V_{i,j}$ for
the various exchange diagrams can be found in Ref. \cite{Holz}. The
dependence on the starting energy $z$ is removed via the prescriptions
given in Eq. (4.7) of Ref. \cite{Reu}. 
Note that the potential matrix elements $V_{i,j}$ are derived by
assuming isospin symmetry. However, the Lippmann-Schwinger equation
(\ref{tmat}) is solved in particle space using the proper 
physical masses of the baryons for the various $\Sigma N$ channels. 
Furthermore, in the charged channels the Coulomb potential is taken
into account. Since we solve the Lippmann-Schwinger equation in 
momentum space this is done by means of the Vincent-Phatak
method \cite{Holz,Phatak}. 

\subsection{Hyperon-nucleon observables}

In Fig.~\ref{cross} we compare the integrated cross sections obtained from 
the new $YN$ potential (solid curves) with the $YN \rightarrow Y'N$ scattering data.
Obviously, a good reproduction of the empirical data \cite{Alex,Sechi,Kadyk,Eisele,Engel} 
is achieved.  Also shown are results from the original J\"ulich $YN$ model~A
\cite{Holz} (dash-dotted curves).
The main qualitative differences between the two models appear in the
$\Lambda p \rightarrow \Lambda p$ channel, for which the J\"ulich model
\cite{Holz} (with standard $\sigma$ and $\rho$ exchange) predicts a broad
shoulder at $p_{lab} \approx$ 350 MeV/c.
This structure, which is not supported by the available experimental
evidence, is due to a bound state in the $^1S_0$ partial wave of the
$\Sigma N$ channel. It is not present in the new model anymore. (We should
say, however, that the new model has a bound state, too. But with a binding
energy of about 400 MeV below the $\Lambda N$ threshold it is located 
completely outside of the physical region. One could speculate, of course, 
that this bound state is a manifestation of the Pauli forbidden $(11)_s$ 
state at the quark level \cite{FujiwaraP}.)
Furthermore, the cusp structure at the opening of the $\Sigma N$ 
threshold is much less pronounced in the new model. In the old model 
this structure was primarily caused by a large amplitude in the 
tensor-coupled $^3S_1-^3D_1$ partial wave of the $\Lambda N$ -- 
$\Sigma N$ transition. This amplitude is now much smaller. As a
consequence also the transition cross section for 
$\Sigma^- p \to \Lambda n$ is now somewhat smaller, though still in
line with the empirical informations. In the $\Sigma^- p$ channel the
new model yields a stronger energy dependence of the reaction cross 
section as it is favoured by
the available cross-section data. In the other two measured
reaction channels the agreement with the data is equally good, if not 
better, for the new model.
 
Note that the $\Sigma^+p$ and $\Sigma^-p$ elastic cross
sections are not ``true'' total cross sections. 
The cross sections that were measured are defined as~\cite{Eisele}
\begin{equation}
  \sigma=\frac{2}{\cos\theta_{\rm max}-\cos\theta_{\rm min}}
         \int_{\cos \theta_{\rm min}}^{\cos \theta_{\rm max}}
         \frac{d\sigma(\theta)}{d\cos\theta}d\cos\theta,
\end{equation}
with typical values $-0.2$ to $-0.5$ for $\cos\theta_{\rm min}$ and
$0.3$ to $0.5$ for $\cos\theta_{\rm max}$. In order to stay as close
as possible to the plotted experimental data, the theoretical curves
in Figs.~\ref{diff}(c) and (d) have been calculated with
$\cos\theta_{\rm min}=-0.5$ and $\cos\theta_{\rm max}=0.5$.

Cross sections at somewhat higher energies are presented 
in Fig.~\ref{cross2}. Note that the data shown in this figure have 
not been taken into account in the fitting process and therefore the 
results are genuine predictions of the model. 
Also here the agreement with the data is satisfactory.   
 
The differential $YN$ scattering cross sections presented in 
Fig.~\ref{diff} are likewise genuine predictions of our $YN$ model.
We want to point out that the empirical information in those 
figures comes from data 
taken from a finite momentum interval, e.g. 160 $ < p_{Lab} < $ 180 MeV/c
for the $\Sigma^+p$ channel \cite{Alex}, whereas the
calculations were performed for the central value of that momentum
interval as it is given in the various plots. 
Note also that the original $YN$ model of the J\"ulich group was
fitted to the data without including the Coulomb interaction (and
without taking into account the mass splitting between $\Sigma^-$, 
$\Sigma^0$, and $\Sigma^+$).  
Thus, the corresponding results presented in Fig.~\ref{diff} do not 
show the strong forward peak caused by the Coulomb amplitude in the
charged channels.  

Evidently, also the data on differential cross sections are 
rather well reproduced by our new $YN$ model. In comparison to
the results of the original J\"ulich model one can say that the
angular dependence in the $\Sigma^- p$ channel is now much better
described and it seems to be more in line with the trend of the 
angular dependence exhibited by the data in the 
$\Sigma^- p \to \Lambda n$ channel too. 

The dashed curves in Figs.~\ref{cross}, \ref{cross2} and 
\ref{diff} are results from an alternative model where the
contributions from the disputed $\kappa$(1000) meson 
have been replaced by a contact interaction. Obviously there is 
practically no sensitivity to the concrete range of the contribution
in the scalar channel with isospin 1/2 -- besides that it has to be 
of fairly short range. In this context we want to mention that we 
could achieve a comparable description of the data even with a 
$\kappa$ mass as low as 800 MeV \cite{Aitala}.

For exploring the differences between the original J\"ulich $YN$ model 
and our new model in more detail we present in Figs.~\ref{pol}, 
\ref{pols} further observables where, however, no data are available. 
Fig.~\ref{pol} contains differential cross sections, polarizations and the
depolarization parameter $D_{nn}$ (definition and explicit expressions 
for those observables can be found in the appendix B of Ref.~\cite{Reu})
for the $\Lambda N$ channel. We present predictions at two energies, 
one ($p_{lab}$ = 150 MeV/c) close to the $\Lambda N$ threshold and
one ($p_{lab}$ = 600 MeV/c) close to (but below) the $\Sigma N$ threshold. 
The results at the higher energy reveal that the new model differs drastically 
from the old one. The differential cross section in the new model is 
strongly forward-peaked whereas the one of the old models peaks in forward
and backward direction. The polarization and $D_{nn}$ have even different
signs. The observables at the lower energy are still dominated by the 
$S$ waves and therefore exhibit only minor differences. But one can see from
the differential cross section that the onset of higher partial waves 
occurs earlier for the new $YN$ model. 
  
Similarly striking differences are present also in the predictions for other
differential observables though we refrain from showing them here. 

For the various $\Sigma N$ channels we present predictions for the polarization
and the depolarization parameter $D_{nn}$ at $p_{lab}$ = 500 MeV/c, cf. 
Fig.~\ref{pols}. Also here one can see that, in general, there are large
differences between the results of the old and the new model. 

\subsection{Low energy parameters and phase shifts}

For the computation of the low energy parameters and phase shifts we 
omit the Coulomb interaction and ignore the mass differences between
the $\Sigma$'s and proton and neutron so that we can solve the 
Lippmann-Schwinger equation in isospin basis. This allows us to
present also results for the $\Sigma N$ system in the $I = 1/2$ 
channel. The $YN$ S-wave low energy parameters are listed in
Table \ref{Effr} while phase shifts for selected partial waves
are shown in Figs.~\ref{phases} and \ref{phases1}.

\begin{table}[ht]
\caption{$YN$ low energy parameters in the $^1S_0$ and $^3S_1$ partial waves
derived from our new model (J04) together with the corresponding results of the 
J\"ulich model A \protect\cite{Holz}. J04c refers to results of an alternative
model where $\kappa$ exchange is replaced by a contact term, cf. text. 
}
\label{Effr}
\begin{center}
\begin{tabular}{|c|c|cccc|}
\hline
 Channel & Model & $a_s(fm)$ & $r_s(fm)$ & $a_t(fm)$ & $r_t(fm)$ \\
\hline
 $\Lambda N$  & J04 & $-$2.56 & 2.75 & $-$1.66 & 2.93 \\
              & J04c & $-$2.66 & 2.67 & $-$1.57 & 3.08 \\
              & A [1] & $-$1.56  & 1.43 & $-$1.59 & 3.16 \\ 
& & & & & \\
 $\Sigma N (I=1/2)$ & J04 & 0.90$-i$0.13 & $-$4.38$-i$2.07 & $-$3.83$-i$3.01 & 2.79$-i$0.57  \\
              & J04c & 0.90$-i$0.13 &$-$4.29$-i$2.05 & $-$3.63$-i$3.09 & 2.78$-i$0.60  \\
              & A [1] & 1.42$-i$0.08 & $-$0.49$-i$0.27 & 2.47$-i$3.74  & 1.61$-i$0.64 \\ 
& & & & & \\
 $\Sigma N (I=3/2)$  & J04 & $-$4.71 & 3.31 & 0.29 & $-$11.54 \\
              & J04c & $-$4.58 & 3.32 & 0.28 & $-$11.63  \\
              & A [1] & $-$2.26 & 5.22  & $-$0.76  & 0.79  \\ 
\hline
\end{tabular} \end{center}
\end{table}

\begin{figure}[ht]
\vglue 2cm
\includegraphics{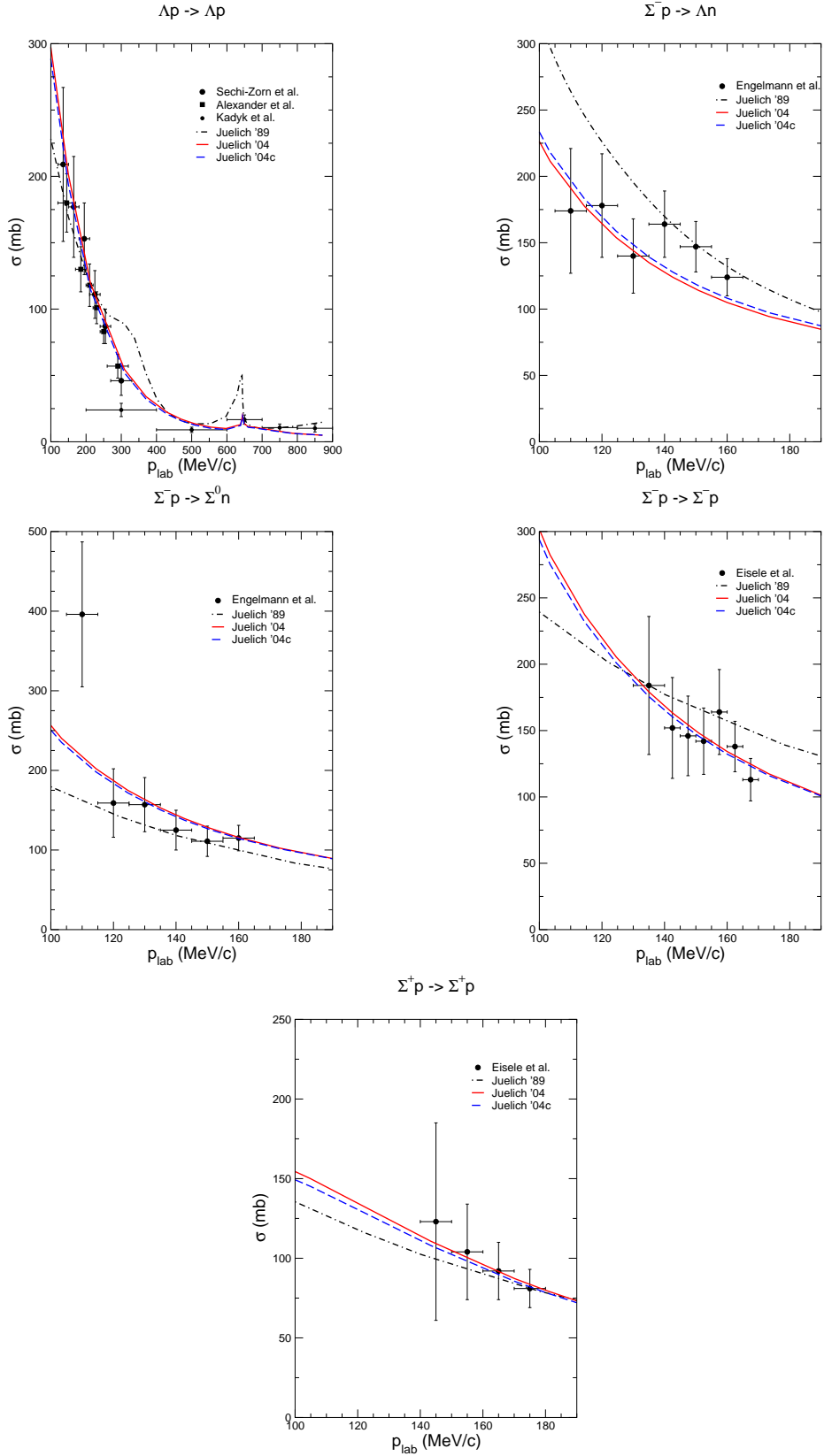} 
\vskip 20cm
\caption{Total $YN$ scattering cross sections as a function of the
        laboratory momentum.
        The solid lines are results of the new $YN$ model, based on
        correlated $\pi\pi$ and $K\bar K$ exchange, while the dash-dotted
        lines are results of the J\"ulich $YN$ model A \protect\cite{Holz}.
        The dashed lines are results of an alternative model where $\kappa$
        exchange is replaced by a contact term, cf. text. 
        The data are from Ref. \protect\cite{Alex,Sechi,Kadyk,Eisele,Engel}.
}
\label{cross}
\end{figure}

\begin{figure}[ht]
\vglue 2cm
\includegraphics{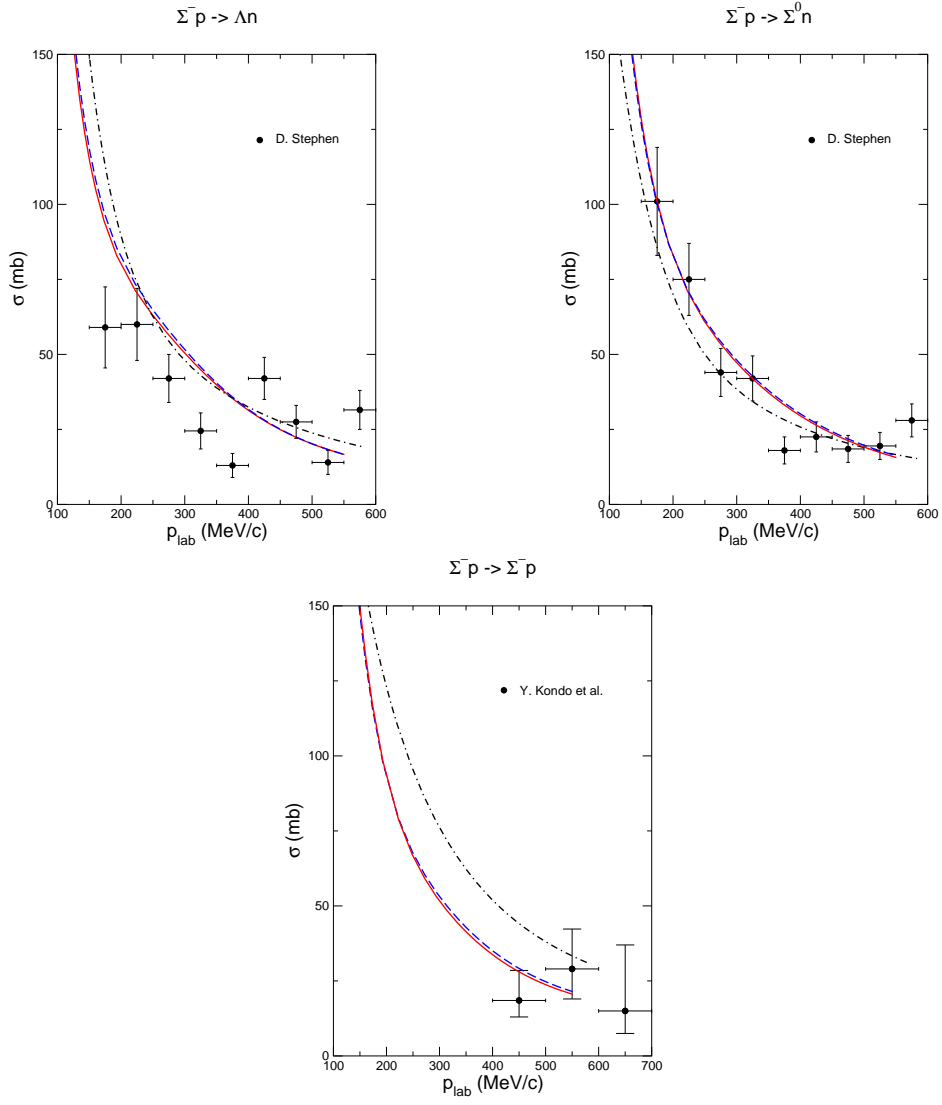} 
\vskip 13cm
\caption{Total $YN$ scattering cross sections as a function of
        the laboratory momentum. Comparison with data at higher energies.
        Same description of curves as in Fig.~\ref{cross}.
        The data are from Ref. \protect\cite{Stephen,Kondo}.}
\label{cross2}
\end{figure}

\begin{figure}[t]
\vglue 2cm
\includegraphics{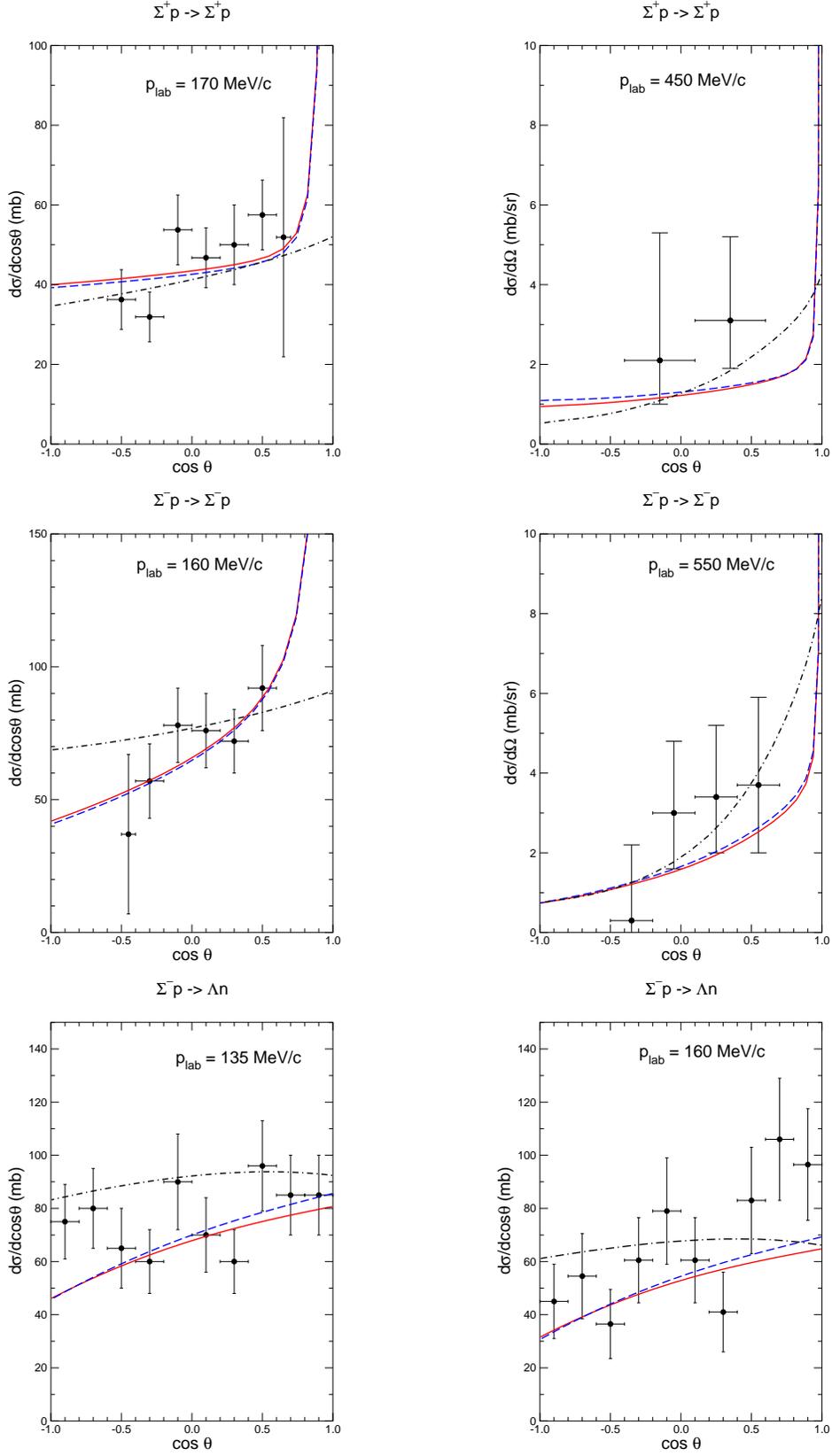} 
\vskip 20cm
\caption{Differential $YN$ scattering cross sections at selected 
        laboratory momenta where data are available. 
        Same description of curves as in Fig.~\ref{cross}.
        The data are from Ref. \protect\cite{Eisele,Engel,Kondo,Ahn}.}
\label{diff}
\end{figure}

\begin{figure}[t]
\vglue 2cm
\includegraphics{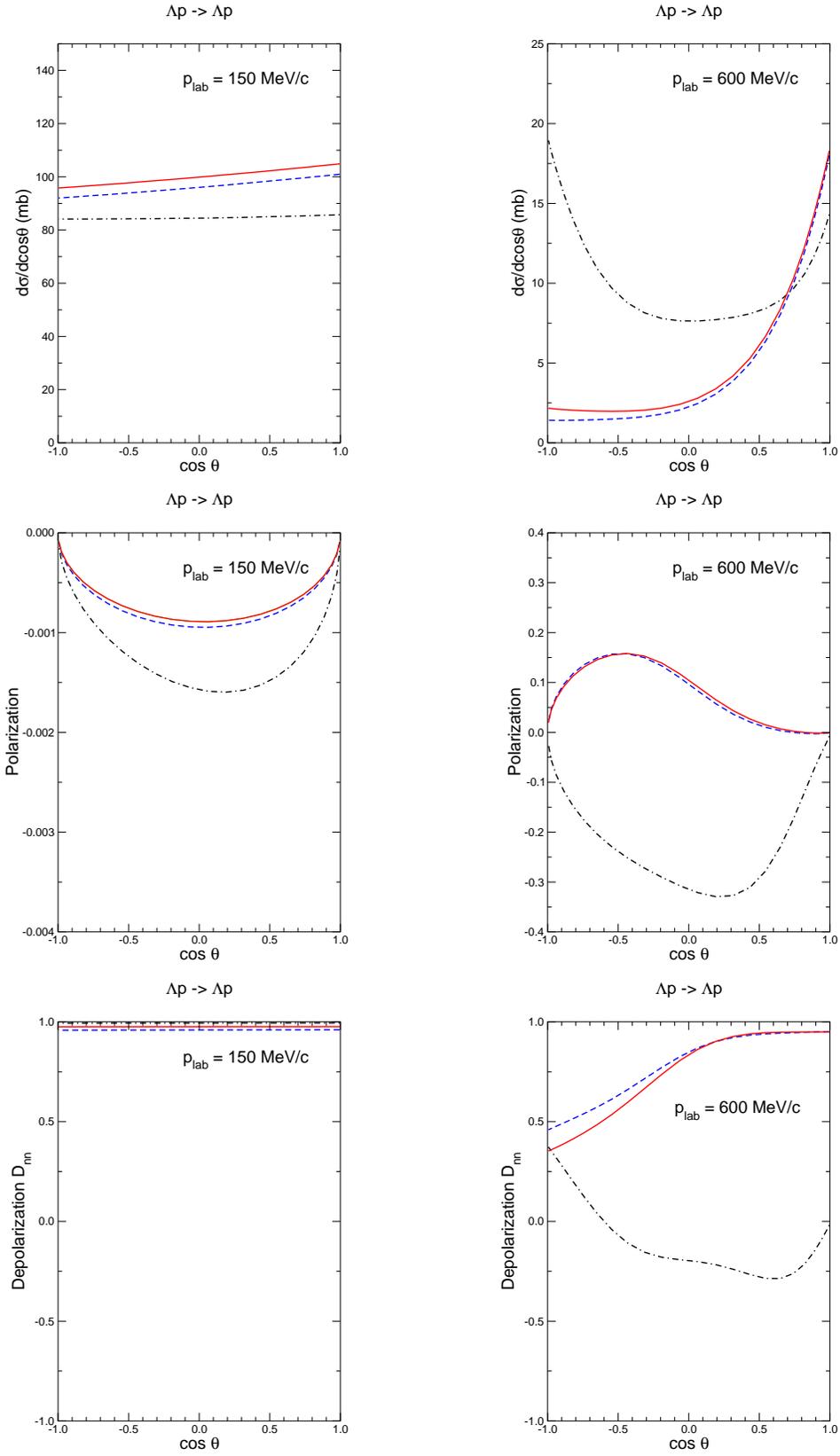} 
\vskip 20cm
\caption{Some differential observables for the reaction $\Lambda N$ 
        at the laboratory momenta 150 and 600 MeV/c. 
        Same description of curves as in Fig.~\ref{cross}.
}
\label{pol}
\end{figure}

\begin{figure}[t]
\vglue 2cm
\includegraphics{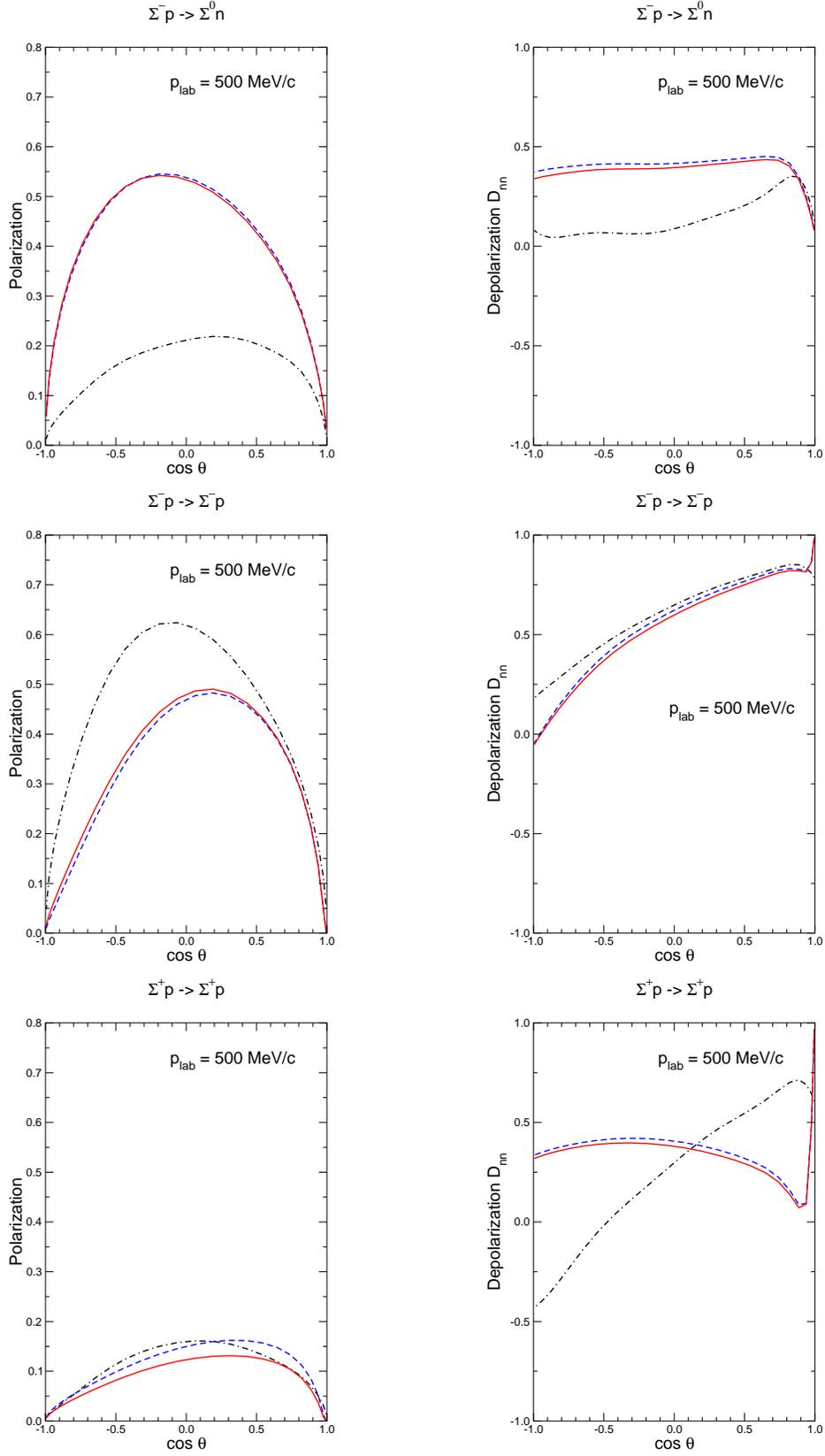} 
\vskip 20cm
\caption{Differential cross sections and polarizations for the reactions 
        $\Sigma N$ at the laboratory momentum 500 MeV/c. 
        Same description of curves as in Fig.~\ref{cross}.
}
\label{pols}
\end{figure}

\begin{figure}[t]
\vglue 2cm
\includegraphics{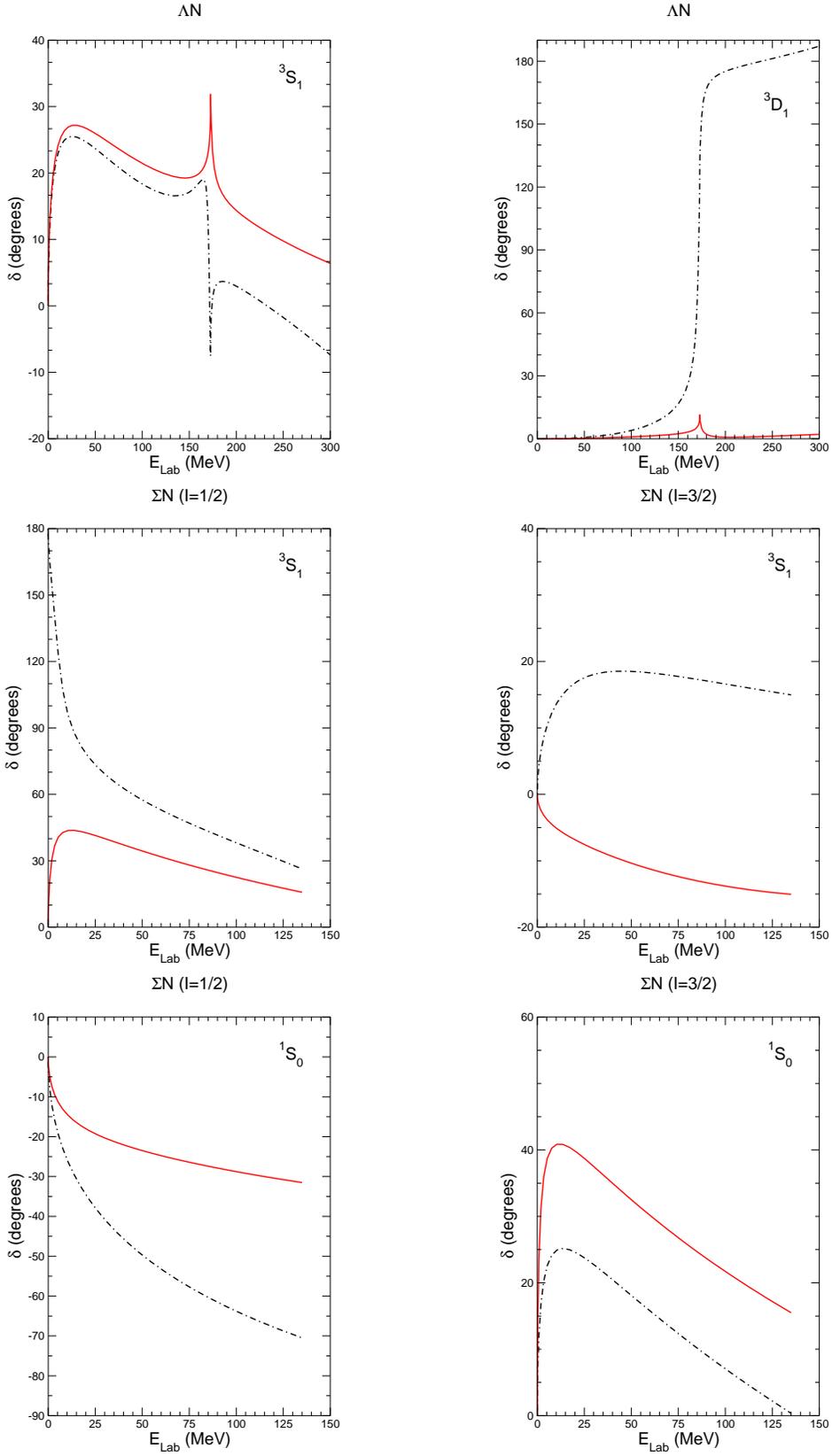} 
\vskip 20cm
\caption{Phase shifts for selected $\Lambda N$ and $\Sigma N$ 
         $S$ and $D$ waves as a function of the laboratory energy. 
        Same description of curves as in Fig.~\ref{cross}.
}
\label{phases}
\end{figure}

\begin{figure}[t]
\vglue 2cm
\includegraphics{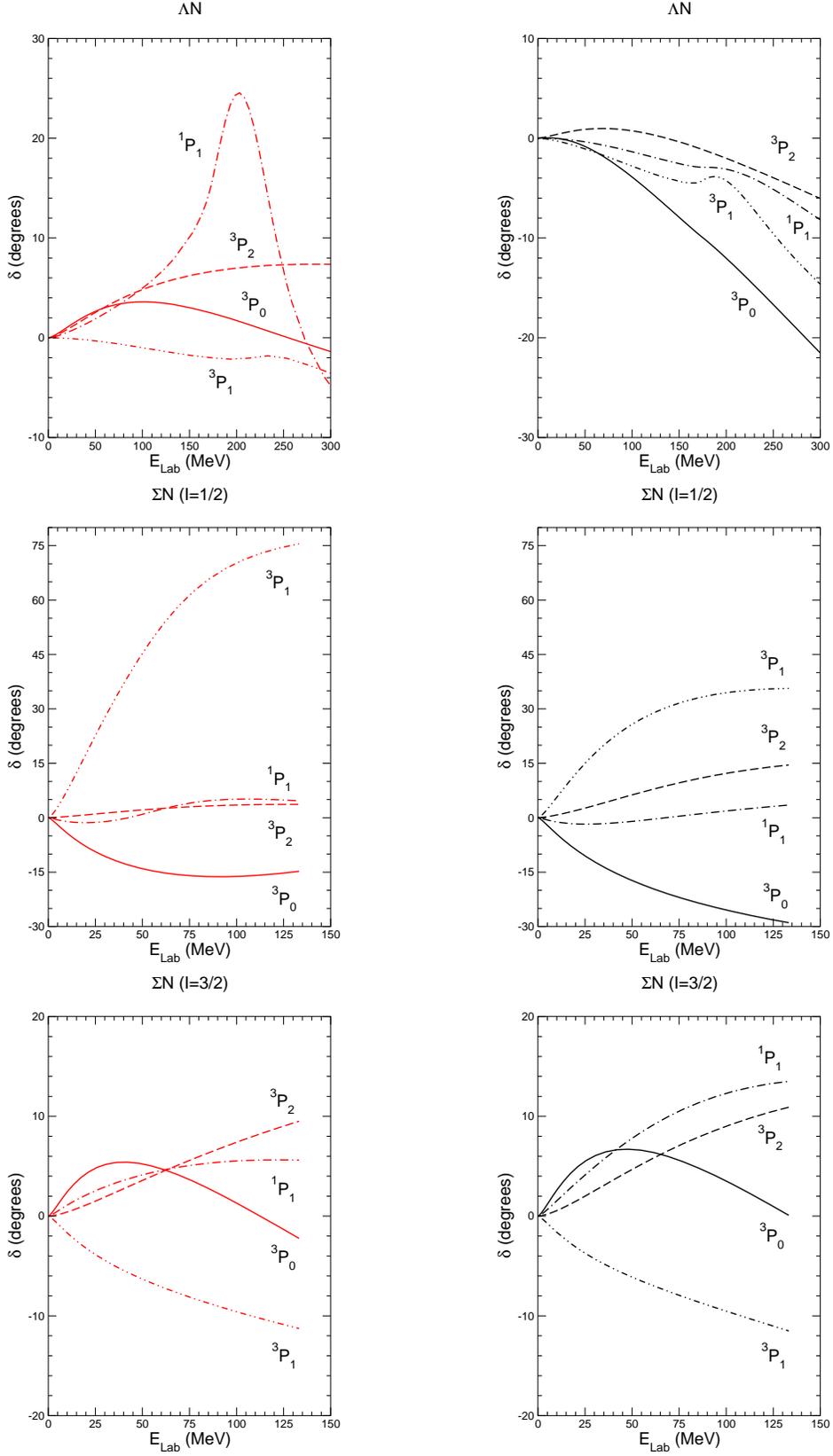} 
\vskip 20cm
\caption{Phase shifts for $\Lambda N$ and $\Sigma N$ 
        $P$ waves as a function of the laboratory energy. 
        The left panel shows results of our new $YN$ model 
        and the right panel those of the original J\"ulich 
        $YN$ model \cite{Holz}. 
}
\label{phases1}
\end{figure}

{}From Table \ref{Effr} one can see that the scattering lengths in the
$^3S_1$ $\Lambda N$ partial wave ($a_t$) are of similar magnitude for
the old and new $YN$ models, but in the $^1S_0$ state ($a_s$) the new
model yields a significantly larger value. The stronger $^1S_0$ 
component of the new model is reflected in the larger $\Lambda p$
cross section near threshold, cf. Fig.~\ref{cross}, and it is 
expected to provide sufficient strength in order to support a
bound hypertriton state \cite{Nogga}. Indeed recent $YN$ models like NSC97f of the 
Nijmegen group \cite{NijV} or the Ehime model 00A \cite{Ehime},
that apparently lead to a bound hypertriton \cite{Miya3,Ehime}, predict 
singlet scattering lengths that are very similar to that of our new model. 

In this context we want to mention that the static version of the old
J\"ulich $YN$ model \cite{Reu} did not support a hypertriton bound
state \cite{Miya1}. However, in that model both the $^1S_0$ as well as
the $^3S_1$ $\Lambda N$ scattering lengths are considerably smaller \cite{Reu}
than in our new $YN$ model.

The scattering lengths and effective ranges for $\Sigma N$ with $I= 1/2$ 
are complex because this channel is coupled to the $\Lambda N$ system.
In the singlet case the scattering lengths are comparable for the two models
whereas in the triplet case they even have opposite signs. 
We want to emphasize, however, that in both models the latter partial wave is 
attractive. But in the original J\"ulich model the attraction is so strong
that there is a near-threshold quasibound state in the $\Sigma N$ channel 
that causes the real part of $a_t$ to be positive - like in case of the 
corresponding $NN$ partial wave and the deuteron. Let us mention that 
practically the same situation occurs in the Nijmegen model NSC97f,
whose pole structure has been investigated and thoroughly discussed in
Ref. \cite{Yamamura}. As a consequence of the near-threshold pole both
these models yield a very pronounced cusp-like structure in the $\Lambda p$
cross section at the opening of the $\Sigma N$ channel, cf. Fig.~\ref{cross}
and Fig.~2 in Ref. \cite{NijV}, respectively. In our new $YN$ model, on the
other hand, the cusp at the $\Sigma N$ threshold is much less pronounced.  
Note that the $^1S_0$ partial wave is attractive too. As already mentioned above,
in the original J\"ulich model there is a bound state in the $\Sigma N$ channel
-- as evidenced by the broad bump in the $\Lambda p$ cross section around
$p_{lab} \approx$ 350 MeV/c. And the new $YN$ model has also a bound state
which is located, however, around 400 MeV below the $\Lambda N$ threshold and 
therefore completely outside of the physically relevant region. 

Let us finally come to the $\Sigma N$ channel with $I= 3/2$. Here we see that
the singlet scattering length of the new model is about twice as large as the
one of the original J\"ulich model. Note that a comparably large singlet
scattering length is also predicted by all of the $YN$ models presented in 
Ref. \cite{NijV}. The scattering lengths for $^3S_1$ are small in both cases,
but of opposite sign. Now, however, it is indeed so that our new $YN$ model
is repulsive in this partial wave whereas the old model is attractive. It is
interesting that basically all available $YN$ models predict rather small
values for the spin-triplet scattering length of the $\Sigma N$ $I= 3/2$
channel \cite{Holz,NijIII,NijIV,NijV,Fujiwara}, though there is no general
trend as far as the sign is concerned. We also observe an unnaturally large 
value for the triplet effective range, which is clearly related to the strong
suppression of the corresponding scattering length. Such a scenario will
require special attention when this channel is considered in effective
field theory (for further discussion, see Sec.~\ref{sec:sum}).
 
Predictions for $\Lambda N$ and $\Sigma N$ phase shifts for selected
$S$- and $D$-waves are shown in Fig.~\ref{phases} and those for 
$P$-waves can be found in Fig.~\ref{phases1}. 
The $\Sigma N$ $S$-wave phase shifts reflect the features that we already
discussed in the context of the scattering lengths. For example one can see
that the phase shift for the $^3S_1$ $I=1/2$ state starts at 180$^0$ for
the original J\"ulich model, as it is expected for a partial wave where
a bound state is present. For the $I=3/2$ state the corresponding phase
is positive, reflecting an attractive interaction, whereas the phase shift
resulting from the new $YN$ model is negative. Note that the phases 
for $^1S_0$ and $I=1/2$ should both start at 180$^0$ because, as mentioned 
above, there is a bound state in both models. 

The opening of the $\Sigma N$ channel at around $E_{Lab} \approx$ 170 MeV 
is cleary reflected in the $^3S_1$ phase shift of the $\Lambda N$ system.  
But its effect on the $^3D_1$ phase shift is even more striking where, 
for the old J\"ulich model, the phase even goes through 90 degrees. 
In fact, the resonance-like behaviour in that partial wave is predominantly 
responsible for the strong enhancement of the $\Lambda N$ cross section in the
vicinity of the $\Sigma N$ threshold, cf. Fig. \ref{cross}. In addition, 
the transition amplitude $^3D_1 (\Lambda N) \leftrightarrow ^3S_1 (\Sigma N)$  
provides a significant contribution to the $\Sigma^-p \to \Lambda n$ cross
section. In the new model the $^3D_1$ phase shift of the $\Lambda N$ system
is much smaller. Accordingly, the cusp-like structure at the $\Sigma N$ threshold
is much less pronounced and the $\Sigma^-p \to \Lambda n$ cross
section is somewhat reduced in this model, as can be seen in Fig. \ref{cross}.

The predictions for the $P$ waves (Fig. \ref{phases1}) show a varying picture.
In the $\Lambda N$ system most of the phases are now attractive whereas they
are mostly repulsive for the old model. This concerns in particular the 
$^1P_1$ amplitude, which is fairly large in the new model, but also the 
$^3P_0$ partial wave. 
In the $I=3/2$ channel of the $\Sigma N$ system the results of the two
models are qualitatively rather similar. To some extent this is also the
case for the $I=1/2$ channel though here the $^3P_1$ amplitude of the new
$YN$ model is significantly larger than the one of the old J\"ulich model. 
Indeed the simultaneous enhancement in the $^3P_1$ ($\Sigma N$) and 
$^1P_1$ ($\Lambda N$) phase shifts is caused by a stronger antisymmetric 
spin-orbit force between the $\Lambda N$ and $\Sigma N$ channels 
in the new model. The increase is primarily due to the $\rho$ exchange 
contribution whose strength for the $\Lambda N \to \Sigma N$ transition, 
fixed from correlated $\pi\pi - K\bar K$ exchange, is
about twice as large as what was used in the old J\"ulich model, 
cf. Table 11 of Ref. \cite{REUBER}. 
In this context let us mention that some other $YN$ models 
exhibit a similarly strong coupling between those partial waves
and channels \cite{FujiwaraA}.

\section{Summary and outlook}
\label{sec:sum}

We have presented a meson-exchange model of the $YN$ interaction where 
-- as the main new feature -- the contributions both in the
scalar-isoscalar  ($\sigma$) and the vector-isovector ($\rho$)
channels are constrained by a microscopic model of correlated
$\pi\pi$ and $K\bar K$ exchange. 

An essential part of baryon-baryon interactions is the strong
medium-range attraction, which in one-boson-exchange models is
parameterized by exchange of a fictitious scalar-isoscalar meson
with mass around 500 MeV.
In extended meson exchange models this part is naturally generated
by two-pion exchange contributions.
As well as uncorrelated two-pion exchange, correlated contributions
must be included in which the exchanged pions interact during their
exchange; these terms in fact provide the main contribution to the
intermediate-range interaction.

As kaon exchange is an essential part of hyperon-nucleon interactions
a simultaneous investigation of correlated $\pi\pi$ and $K\anti{K}$
exchanges is clearly necessary.
In Ref.~\cite{REUBER} the correlated $\pi\pi$ and $K\anti{K}$ exchange
contributions in various baryon-baryon channels have therefore been
investigated within a microscopic model for the transition amplitudes
of the baryon-antibaryon system ($B\anti{B'}$) into $\pi\pi$ and
$K\anti{K}$ for energies below the $B\anti{B'}$ threshold.
The correlations between the two mesons have been taken into account
by means of $\pi\pi-K\anti{K}$ amplitudes, determined in the field
theoretical framework of Refs.~\cite{Lohse,Janssen,Schutz}, which provide an
excellent description of empirical $\pi\pi$ data up to 1.3 GeV.
With the help of unitarity and dispersion-theoretical methods, the
baryon-baryon amplitudes for correlated $\pi\pi$ and $K\anti{K}$
exchange in the $J^P=0^+$ ($\sigma$) and $J^P=1^-$ ($\rho$)
$t$-channels have then been determined.
With this model it is possible to reliably take into
account correlated $\pi\pi$ and $K\anti{K}$ exchange in both the
$\sigma$ and $\rho$ channels for various baryon-baryon reactions.
Given the strong constraints on $\sigma$ as well as $\rho$ exchange
from correlated $\pi\pi$ exchange, a more sound microscopic model
for the $YN$ interaction can hence now be constructed.

Besides contributions from correlated $\pi\pi$ and $K\anti{K}$ exchange 
the present model incorporates also the standard one-boson exchanges
of the lowest pseudoscalar and vector meson multiplets with coupling
constants fixed by SU(6) symmetry relations. Thus, in the
present model the long- and intermediate-range part of the $YN$ 
interaction is completely determined -- either by SU(6) constraints 
or by correlated $\pi\pi$ and $K\bar K$ exchange. 

In addition there are some short-ranged ingredients. 
First of all, the contribution from the $a_0(980)$ meson is taken into
account. Secondly, we consider the exchange of a strange scalar meson, the
$\kappa$, with mass $\sim 1000$~MeV. (Note that these pieces 
were not taken into account in the earlier $YN$ models of the J\"ulich
group \cite{Holz,Reu}.) These short-ranged contributions are also viewed as
being due to correlated meson-meson exchanges but in practice they are
parametrized phenomelogically in terms of one-boson-exchange 
contributions in the corresponding spin-isospin channels. In particular, 
no SU(3) relations are imposed on the short-range part. This 
assumption is based on our observation that the contributions in the 
$\rho$ exchange channel as they result from correlated $\pi\pi$ and 
$K\bar{K}$ no longer fulfill SU(3) relations, but it also acknowledges 
the fact that at present there is no general agreement about who are
the actual members of the lowest-lying scalar meson SU(3) multiplet.

The new $YN$ model provides a rather satisfactory reproduction of the 
available $YN$ data. It describes not only the integrated cross
sections for $\Lambda p$ and the various $\Sigma N$ channels but
also the few available data on differential cross sections, even 
though the latter were not included in the fitting procedure. 
We see that as an indication that the data are compatible with 
the assumption of SU(6) symmetry for the pseudoscalar sector of
our $YN$ model. 

As the main qualitative difference between the old $YN$ J\"ulich model
\cite{Holz} (with standard $\sigma$ and $\rho$ exchange) 
we want to mention that the broad shoulder at $p_{lab} \approx$ 
350 MeV/c in the $\Lambda p \rightarrow \Lambda p$ channel, 
predicted by that model but not seen in the experiments,
is no longer present in the new model. But, as a more detailed
comparison revealed, there are also striking differences between these
two models in the predictions for the individual partial waves. 
For example, in the new model the triplet $S$ wave in the $I=3/2$ 
channel of the $\Sigma N$ system is repulsive  
and some of the $P$-wave amplitudes are significantly larger. 
Thus, it will be interesting to see the performance of the new
$YN$ interaction model in applications to few- and many-body
systems involving hyperons \cite{Nogga}. 

This study also paves the way for a systematic investigation in the
framework of effective field theory, see \cite{Henk}. In such a framework,
pion- and kaon exchange supplemented by four-baryon contact interactions
(these encode the contributions from the exchange of heavier mesons not
linked to chiral symmtery) is considered to generate a potential based 
on the power counting rules. It remains to be seen how well such a more
systematic approach can indeed describe the data and what conclusions can
be drawn about three-baryon forces that naturally arise in such a framework.

\section*{Acknowledgements}
We thank J.~Speth and W.~Melnitchouk for collaboration during the early stages
of this investigation. We also thank A. Nogga for a careful reading of our
manuscript. This research is part of the EU Integrated Infrastructure Initiative 
Hadron Physics Project under contract number RII3-CT-2004-506078. The work was 
supported in part by DFG through funds provided to the special research grant
TR-16 ``Subnuclear Structure of Matter''.

\def\Nucl{Nucl.\ }
\def\Phys{Phys.\ }
\def\Rev{Rev.\ }
\def\Lett{Lett.\ }
\def\PL{\Phys\Lett}
\def\PLB{\Phys\Lett B }
\def\NP{\Nucl\Phys}
\def\NPA{\Nucl\Phys A }
\def\NPB{\Nucl\Phys B }
\def\NPBS{\Nucl\Phys (Proc.\ Suppl.\ )B }
\def\PR{\Phys\Rev}
\def\PRL{\Phys\Rev\Lett}
\def\PRC{\Phys\Rev C }
\def\PRD{\Phys\Rev D }
\def\RMP{\Rev  Mod.\ \Phys}
\def\ZP{Z.\ \Phys}
\def\ZPA{Z.\ \Phys A }
\def\ZPC{Z.\ \Phys C }
\def\AOP{Ann.\ \Phys}
\def\PRep{\Phys Rep.\ }
\def\ANP{Adv.\ in \Nucl\Phys Vol.\ }
\def\PTP{Prog.\ Theor.\ \Phys}
\def\PTPS{Prog.\ Theor.\ \Phys Suppl.\ }
\def\PL{\Phys \Lett}
\def\JPF{J.\ Physique}
\def\FBSS{Few--Body Systems, Suppl.\ }
\def\IJMP{Int.\ J.\ Mod.\ \Phys A}
\def\NuCi{Nuovo Cimento~}

%%%%%%%%%%%%%%%%%%%%%%%%%%%%%%%%%%%%%%%%%%%%%%%%%%%%%%%%%%%%%%%%%%%%%%%%%%
\newpage 

\section*{Figures}

\vspace{1.0cm}

\vspace{0.5cm}

\vspace{0.5cm}

%%%%%%%%%%%%%%%%%%%%%%%%%%%%%%%%%%%%%%%%%%%%%%%%%%%%%%%%%%%%%%%%%%%%%%%%%%
\newpage

\vspace{0.5cm}

%%%%%%%%%%%%%%%%%%%%%%%%%%%%%%%%%%%%%%%%%%%%%%%%%%%%%%%%%%%%%%%%%%%%%%%%%%
\newpage 

%%%%%%%%%%%%%%%%%%%%%%%%%%%%%%%%%%%%%%%%%%%%%%%%%%%%%%%%%%%%%%%%%%%%%%%%%%
\newpage 

%%%%%%%%%%%%%%%%%%%%%%%%%%%%%%%%%%%%%%%%%%%%%%%%%%%%%%%%%%%%%%%%%%%%%%%%%%
\newpage 

%%%%%%%%%%%%%%%%%%%%%%%%%%%%%%%%%%%%%%%%%%%%%%%%%%%%%%%%%%%%%%%%%%%%%%%%%%
\newpage 

%%%%%%%%%%%%%%%%%%%%%%%%%%%%%%%%%%%%%%%%%%%%%%%%%%%%%%%%%%%%%%%%%%%%%%%%%%
\newpage 

%%%%%%%%%%%%%%%%%%%%%%%%%%%%%%%%%%%%%%%%%%%%%%%%%%%%%%%%%%%%%%%%%%%%%%%%%%
\newpage 

%%%%%%%%%%%%%%%%%%%%%%%%%%%%%%%%%%%%%%%%%%%%%%%%%%%%%%%%%%%%%%%%%%%%%%%%%%
\newpage 

%%%%%%%%%%%%%%%%%%%%%%%%%%%%%%%%%%%%%%%%%%%%%%%%%%%%%%%%%%%%%%%%%%%%%%%%%%
\newpage 

\end{document}